\documentclass[article,10pt,nofootinbib,notitlepage,twocolumn]{revtex4-2}
\usepackage{graphicx}
\usepackage{amsmath}
\usepackage{amssymb}
\usepackage{dcolumn}
\usepackage{bm}
\usepackage{color}
\usepackage{dsfont}
\usepackage{feynmp}
\usepackage{marvosym}
\usepackage{phaistos}
\usepackage[hidelinks=true]{hyperref}

\newcommand \beq{\begin{eqnarray}}
\newcommand \eeq{\end{eqnarray}}

\begin{document}
\unitlength=1mm
\allowdisplaybreaks

\title{Deconfinement transition within the Curci-Ferrari model\\ -- Renormalization scale and scheme dependences}

\author{V. Tomas \surname{Mari Surkau}}
\affiliation{Centre de Physique Th\'eorique, CNRS, Ecole polytechnique, IP Paris, F-91128 Palaiseau, France.}

\author{Urko Reinosa}
\affiliation{Centre de Physique Th\'eorique, CNRS, Ecole polytechnique, IP Paris, F-91128 Palaiseau, France.}

\date{\today}

\begin{abstract}
We analyze the confinement/deconfinement transition of pure Yang-Mills theories within the framework of the center-symmetric Landau gauge supplemented by a Curci-Ferrari mass term that models the effect of the associated Gribov copies in the infrared. In addition to providing details for earlier one-loop calculations in that framework, we explore how the results depend on the renormalization scale and/or on the renormalization scheme.  We find that the predicted values for the transition temperatures of SU($2$) and SU($3$) Yang-Mills theories are similar in both schemes and are little sensitive to the renormalization scale $\mu$ over a wide range of values including the standard range $\smash{\mu\in[\pi T,4\pi T]}$. These values are also close both to those obtained from a minimal sensitivity principle and to those of lattice simulations, especially in the SU($3$) case. These results further confirm the good behavior of perturbative calculations within the Curci-Ferrari model and support the adequacy of the latter as an effective description of Yang-Mills theories in the infrared. We perform a similar analysis for the spinodal temperatures in the SU($3$) case and for the Polyakov loop, the order parameter associated to the breaking of center symmetry.
\end{abstract}

\maketitle

\pagebreak

\section{Introduction}

The Landau gauge is a very popular choice within the corpus of functional approaches to non-Abelian gauge theories \cite{Alkofer:2000wg,Zwanziger:2001kw,Fischer:2003rp,Aguilar:2004sw,Boucaud:2006if,Aguilar:2007ie,Aguilar:2008xm,Boucaud:2008ky,Fischer:2008uz,Rodriguez-Quintero:2010qad,Pawlowski:2003hq,Fischer:2004uk,Cyrol:2016tym,Dupuis:2020fhh}. In particular, this gauge can be implemented within the framework of Monte-Carlo importance sampling techniques and has thus opened the way to an active and fruitful cooperation between continuum and discrete approaches. 

The practical implementation of the Landau gauge is hindered, however, by the Gribov ambiguity \cite{Gribov77} which, although not relevant at high energies, is believed to play a role in the study of the infrared properties of non-Abelian gauge theories. There exist various ways by which one can try to take into account the ambiguity \cite{Zwanziger89,Vandersickel:2012tz,Dudal:2008sp}. Among them, the one based on the Curci-Ferrari (CF) action \cite{Curci:1976bt} is a phenomenological approach \cite{Pelaez:2021tpq} that capitalizes on the results of Landau gauge-fixed lattice simulations obtained over the past decades \cite{Bonnet:2000kw,Bonnet:2001uh,Cucchieri_08b,Bogolubsky09,Bornyakov09,Iritani:2009mp,Maas:2011se,Oliveira:2012eh,Boucaud:2011ug}. Most notably, in the zero-momentum limit, the lattice gluon propagator saturates to a finite non-zero value. This mass scale is accommodated within the CF model by adding a gluon mass term to the usual, but incomplete,\footnote{Given that it disregards the Gribov copies.} Landau-gauge Faddeev-Popov action. The same lattice simulations also show that, in the pure glue sector, the coupling never becomes too large,\footnote{By coupling, we here mean the rescaled coupling $g^2N/(16\pi^2)$ which is the actual expansion parameter of perturbative calculations in the vacuum.} which points to the idea that, at least for applications to pure YM theories, the CF model can be considered at a perturbative level. This approach has been quite successful in describing YM correlation functions in the vacuum, in good agreement with the lattice data \cite{Tissier:2011ey,Pelaez:2013cpa,Reinosa:2017qtf,Barrios:2020ubx,Barrios:2022hzr}.\footnote{In the QCD case, the infrared coupling that can be extracted from the vertex functions in the matter sector is larger than the one that can be extracted from the vertex functions in the glue sector. This calls for the use of non-perturbative methods. Interestingly enough, however, this hierarchy in the infrared couplings allows one to devise a non-perturbative expansion with good control on the error \cite{Pelaez:2017bhh,Pelaez:2020ups}.}

When extending these considerations to the finite temperature case, one needs to pay special attention to the center symmetry which controls the confinement/deconfinement transition in the case of pure Yang-Mills theories. Unfortunately, the Landau gauge is not very efficient in this respect, in particular, in the presence of approximations \cite{vanEgmond:2023lnw}. A better choice consists in upgrading the Landau gauge to the class of background Landau gauges \cite{Abbott:1980hw,Abbott:1981ke} and to consider the associated background field effective action \cite{Braun:2007bx,Braun:2010cy}. As the latter is center-symmetric, its absolute minima allow one to distinguish between the Wigner-Weyl and Nambu-Goldstone realizations of the symmetry. 

The background field effective action is not a Legendre Transform, however, and the very rationale for extracting physical information from its absolute minima relies on one additional property \cite{vanEgmond:2023lnw} which is only approximately satisfied in practice, thus introducing a potential bias in the results. Recently, to cope with this limitation, we have proposed a different approach. It is also set up within the class of background Landau gauges but it uses a genuine Legendre Transform, computed for a particular class of backgrounds known as center-symmetric. Within these ``center-symmetric Landau gauges'', the effective action is center-symmetric and its minima allow again for the identification of explicitly realized or broken symmetry phases, the major difference being that the very use of the minima does not rely in this case on any extra assumption.

As in the case of the standard Landau gauge, however, the center-symmetric Landau gauges suffer from the Gribov ambiguity and it is necessary to try to take the latter into account in one way or another. One possibility is to extend the CF model in the presence of a center-symmetric background \cite{vanEgmond:2023lnw}. In Ref.~\cite{vanEgmond:2021jyx}, the confinement/deconfinement transition has been studied using this extension to one-loop accuracy, with predictions for the SU($2$) and SU($3$) transition temperatures in pretty good agreement with the results of lattice simulations, especially in the SU($3$) case.

It should be stressed, however, that those results have been obtained within one specific renormalization scheme and for a fixed value of the renormalization scale, the same one that was used to determine the parameters of the CF model from the fits of Landau gauge lattice propagators. A more complete analysis requires assessing how much the results depend on the renormalization scale and/or on the choice of the renormalization scheme. 

In this paper, we combine the results of Ref.~\cite{vanEgmond:2021jyx} with a renormalization group analysis, following the lines of Refs.~\cite{Tissier:2011ey} to test the renormalization scale dependence of the results in two different renormalization schemes. While the analysis of the renormalization scale dependence in a given scheme will allow us to test the internal consistency of perturbative calculations within the CF model, the analysis of the scheme dependence will allow us to explore the adequacy of that model for describing Yang-Mills theories in the infrared. This is because the initializations of the renormalization group flows will be taken from fits to the Landau gauge correlators as simulated on the lattice, that is input external to the CF model itself. We stress that our analysis will apply to the strict one-loop calculation performed in Ref.~\cite{vanEgmond:2021jyx}. Of course, one could also use the RG to improve this calculation in various ways. We leave this task, however, for a subsequent analysis.

The paper is organized as follows. In Sec.~\ref{sec:framework}, we briefly review the framework that will be used in this work, including the class of background Landau gauges, the particular instances known as center-symmetric Landau gauges and the associated Curci-Ferrari model. In Sec.~\ref{sec:potential}, we provide some details on the derivation of the formula for the one-loop effective potential and discuss the symmetries of the latter. Section \ref{sec:renormalization} is devoted to the renormalization of the potential while Sec.~\ref{sec:evaluation} discusses its practical evaluation. Finally, Sec.~\ref{sec:schemes} reviews the two renormalization schemes and the associated RG-flows that will be used in the results section, Sec.~\ref{sec:results}. The appendices gather further technical details. Appendix \ref{app:det} describes the steps that lead to the one-loop expression for the effective potential while Appendix \ref{app:other} provides an alternative practical evaluation of the latter, complementing the one presented in Sec.~\ref{sec:evaluation}. Similarly, Appendix \ref{app:sums} discusses the evaluation of Matsubara sum-integrals, especially in those cases not adapted to the standard contour integration technique, while Appendix \ref{app:eps} deals with the specific question of interchanging the order between the Matsubara summation and the continuum limit in dimensional regularization.

\section{The framework}\label{sec:framework}

\subsection{Background Landau gauges}
We consider the class of {\it background Landau gauges} defined by the condition 
\beq
\bar D_\mu(A_\mu^a-\bar A_\mu^a)=0\,,\label{eq:condition}
\eeq 
where $\bar A_\mu^a$ is a given background gauge-field configuration and $\smash{\bar D_\mu^{ac}\equiv \partial_\mu\delta^{ac}+gf^{abc}\bar A_\mu^b}$ is the adjoint covariant derivative for that background. The associated Euclidean, Faddeev-Popov gauge-fixed action is
\beq\label{eq:action}
S_{\bar A}[A]=\!\!\int_x\!\left[\frac{1}{4}F_{\mu\nu}^aF_{\mu\nu}^a\!+\!\bar D_\mu\bar c^a \! D_\mu c^a\!+\!ih^a\!\bar D_\mu(A_\mu^a-\bar A_\mu^a)\right],\nonumber\\
\eeq
where $\smash{F_{\mu\nu}^a\equiv\partial_\mu A_\nu^a-\partial_\nu A_\mu^a+gf^{abc}A_\mu^bA_\nu^c}$ is the non-Abelian field-strength tensor while $c^a$, $\bar c^a$ and $h^a$ denote the ghost, anti-ghost and Nakanishi-Lautrup fields. Note the presence of $\smash{D_\mu c^a\equiv \partial_\mu c^a+gf^{abc} A_\mu^b c^c}$ which does not involve the background.

Here and in the following, we adopt an Euclidean set-up adapted to the finite-temperature case and we recall that the gauge fields are periodic along the Euclidean time direction with a period equal to the inverse temperature $\smash{\beta\equiv 1/T}$:
\beq
A_\mu^a(\tau+\beta,\vec{x})=A_\mu^a(\tau,\vec{x})\,.\label{eq:periodic}
\eeq
Equivalently, they are defined over the compact time interval $[0,\beta]$ and obey periodic boundary conditions.

The very choice of the class of gauges (\ref{eq:condition}) relies on the fact that the gauge-fixed action (\ref{eq:action}) is invariant under the simultaneous ``gauge'' transformation of $A_\mu^a$ and $\bar A_\mu^a$, that is\footnote{It is implicitly understood that the other fields $\varphi^a$ need to be color-rotated as $\smash{\varphi^U=U\varphi U^\dagger}$, where $\smash{\varphi\equiv\varphi^at^a}$.}
\beq
S_{\bar A^U}[A^U]=S_{\bar A}[A]\,,\label{eq:prop}
\eeq
with
\beq
A^U_\mu= UA_\mu U^\dagger+\frac{i}{g}U\partial_\mu U^\dagger\,,\label{eq:AU}\\
\bar A^U_\mu= U\bar A_\mu U^\dagger+\frac{i}{g}U\partial_\mu U^\dagger\,,\label{eq:AbarU}
\eeq
and where we have defined $\smash{A_\mu\equiv A_\mu^at^a}$ and $\smash{\bar A_\mu\equiv\bar A_\mu^a t^a}$. The identity (\ref{eq:prop}) is known as {\it background ``gauge'' symmetry} and plays a pivotal role in the following. We stress that this relation does not reflect any gauge-invariance property of $S_{\bar A}[A]$ since the gauge is anyway fixed by the choice \eqref{eq:condition}. Rather, Eq.~(\ref{eq:prop}) connects two background gauges with respective backgrounds $\bar{A}$ and $\bar{A}^U$.

\subsection{Center-symmetric Landau gauges}
It is important to stress that only those transformations $U$ in Eqs.~(\ref{eq:AU}) and (\ref{eq:AbarU}) that are themselves periodic in time ought to be considered as genuine gauge transformations, that is transformations that do not alter the state of the system. They are generically written as $U_0$ and form a group denoted ${\cal G}_0$.

There is, however, a larger group ${\cal G}$ of transformations that preserve the periodicity condition (\ref{eq:periodic}), namely those transformations $U$ that are periodic modulo an element of the center of SU(N), that is
\beq
U(\tau+\beta,\vec{x})=e^{i2\pi k/N}U(\tau,\vec{x})\,,\label{eq:k}
\eeq
with $\smash{k=0,1,\dots,N-1}$. The transformations corresponding to $\smash{k\neq 0}$ need to be interpreted as genuine physical transformations that alter the state of the system.\footnote{One could of course question this interpretation by stating that all transformations within ${\cal G}$ are actual gauge transformations. However, this would have annoying consequences at an interpretation level. In particular, the Polyakov loop, see below, would not be an observable. Let us also mention that, within QCD, the only transformations that are symmetries of the Euclidean functional integral representation of the partition function are those within ${\cal G}_0$. Then, it makes sense once again to interpret ${\cal G}_0$ as the group of genuine gauge transformations, the other, physical symmetries being explicitly broken by the presence of dynamical quarks in this case.} They change at least one observable characterizing this state, the {\it Polyakov loop,} which measures the free-energy of a heavy test color charge in the thermal bath of gluons. Because the Polyakov loop is invariant under the action of ${\cal G}_0$, the actual physical symmetry group is the quotient group ${\cal G}/{\cal G}_0$. This group is isomorphic to the center of SU(N) and is, therefore, naturally dubbed as the {\it center-symmetry group.} 

As long as this symmetry is explicitly realized in the system, the Polyakov loop needs to vanish, which makes the free energy of the heavy test quark infinite and which, in turn, is interpreted as the confining phase of the thermal bath of gluons. In contrast, the spontaneous breaking of the symmetry allows the Polyakov loop to acquire a non-zero value, corresponding to a finite free energy and thus to a deconfined phase. This shows that the Polyakov loop plays the role of an order parameter for center symmetry and, as such, allows one to identify the confinement/deconfinement transition within the pure Yang-Mills system. This order parameter is gauge-invariant, that is invariant under the action of ${\cal G}_0$.

In the continuum, however, where the Polyakov loop is not the simplest quantity to evaluate, it might be useful to identify simpler order parameters that probe the same symmetry. This can be done, in particular, by choosing a background $\bar A_c$ such that
\beq
\bar A_c^{U_c}=\bar A_c\,,\label{eq:inv}
\eeq
for a certain transformation $U_c$ obeying the condition (\ref{eq:k}) with $\smash{1\leq k<N}$.\footnote{In what follows, we shall work essentially with $\smash{k=1}$.} We refer to these types of backgrounds as center-invariant and to the associated gauges as center-invariant Landau gauges. Now, because of Eqs.~(\ref{eq:prop}) and (\ref{eq:inv}), the transformation $U_c$ is a symmetry of the gauge-fixed action in the corresponding gauge:\footnote{This identity substantially differs from the more general one (\ref{eq:prop}). Indeed, the latter reflects a connection between two distinct gauges characterized by $\bar A$ and $\bar A^U\neq\bar A$, while the former is a symmetry constraint within a particular gauge characterized by $\bar A_c$.}
\beq
S_{\bar A_c}[A^{U_c}]=S_{\bar A_c}[A]\,.\label{eq:sym}
\eeq
This simple observation allows one to define alternative order parameters for the confinement/deconfinement transition in terms of the $n$-point functions \cite{vanEgmond:2022nuo,vanEgmond:2023lfu}. The simplest example, on which we shall focus in the present work, is that of the one-point function $\langle A\rangle_{\bar A_c}$. This proposal differs from the popular choice based on the use of self-consistent backgrounds $\smash{\bar{A}_s=\langle A\rangle_{\bar A_s}}$ and the associated background effective action. We shall compare the two approaches in Sec.~\ref{subsec: background pot}.

\subsection{Constant, temporal, and Abelian backgrounds}
In practice, it is not necessary to determine all possible center-invariant backgrounds. In this respect, it is convenient to first restrict to constant, temporal, and Abelian backgrounds, that is
\beq
\bar A_\mu(x)=\frac{T}{g}\bar r^j t^j \delta_{\mu0}\,,\label{eq:bg}
\eeq
where the $t^j$ span the diagonal part of the algebra and the $\bar r^j$ are the components of a constant, dimensionless\footnote{The factor $T/g$ has dimension $(d-2)/2$ in dimensional regularization, that is precisely the dimension of $\bar A$. We stress that, so far, $g$ denotes the bare coupling which has dimension $(4-d)/2$. The dimensionless, renormalized coupling will be introduced later when dealing with renormalization.} vector $\bar r$ within $\mathds{R}^{N-1}$. The particular values $\smash{\bar r=\bar r_c}$ that correspond to center-invariant backgrounds can then be determined using the notion of Weyl chambers, see below as well as Ref.~\cite{Reinosa:2020mnx,vanEgmond:2023lfu} for details. In the SU($2$) case, one can for instance take $\smash{\bar r_c=\pi}$, with $\smash{t^j=\sigma^3/2}$, while in the SU($3$) case, one can take $\smash{\bar r_c=(4\pi/3,0)}$, with $\smash{t^j\in\{\lambda^3/2,\lambda^8/2\}}$.

Moreover, with the choice of background (\ref{eq:bg}), it can be shown that the one-point function acquires a similar form
\beq
\langle A_\mu\rangle_{\bar A}=\frac{T}{g} r^j t^j \delta_{\mu0}\,,\label{eq:res}
\eeq
and that within the gauge $\smash{\bar r=\bar r_c}$, and as long as center symmetry is not spontaneously broken, $r$ should also be center-symmetric. Any deviation from this allows one to identify the confinement/deconfinement transition. It is the purpose of this paper to analyze how this happens and how this depends on the renormalization scheme and renormalization scale, complementing the results of Ref.~\cite{vanEgmond:2021jyx}.

\subsection{The Curci-Ferrari model}
Before doing so, however, an important remark is in order. The point is that the Faddeev-Popov action (\ref{eq:action}) is not a {\it bona fide} gauge fixing due to the existence of Gribov copies in the class of gauges (\ref{eq:condition}). Even though these copies are expected not to play a role at high energies and the use of the Faddeev-Popov action is sensible in this case, this is not necessarily so at low energies, in particular regarding the low-temperature phase of the system. It is usually accepted that the Faddeev-Popov action needs to be modified in that case.

Here we do not aim at implementing this modification exactly. Rather, we consider a phenomenological take on that question, based on the Curci-Ferrari model which has shown a surprising ability to capture many infrared properties of Landau gauge YM theory from simple perturbative calculations, see \cite{Pelaez:2021tpq} for a thorough review.

In the background Landau gauges, the CF extension of the FP gauge-fixing (\ref{eq:action}) is provided by
\beq\label{eq:S0}
S_{\bar A}[A]+\int_x\frac{1}{2}m^2(A_\mu^a-\bar A_\mu^a)^2\,,
\eeq
where $m$ is the CF mass parameter whose value can be fixed by fitting the zero-temperature\footnote{In this limit, and due to the explicit factor of $T$ in Eq.~(\ref{eq:bg}), the center-symmetric background vanishes and the background Landau gauge coincides with the Landau gauge.} Landau gauge correlators determined in lattice simulations. The mass term in Eq.~(\ref{eq:S0}) is tailor-made such that the symmetry identity (\ref{eq:prop}) is preserved, and we can still rigorously define an order parameter via the relation \eqref{eq:sym}. For other approaches that phenomenologically include massive gluons, see for instance \cite{Meisinger:2001cq,Dumitru:2012fw}. In these approaches, the modelling is usually done at the level of the thermodynamical potential. One benefit of the Curci-Ferrari approach is that it is formulated, instead, at the level of the classical action, thus allowing for the evaluation of higher order corrections.

\section{The one-loop effective potential}\label{sec:potential}
In what follows, we determine the one-point function $\langle A_\mu\rangle_{\bar A_c}$ within the Curci-Ferrari model in the presence of a center-invariant background $\bar A_c$. To take into account the possibility of spontaneous center-symmetry breaking, one needs to perform the calculation in the presence of an external source $J_\mu$ coupled to $A_\mu$ that breaks explicitly the symmetry (\ref{eq:sym}), and, only then, to analyze what happens as the external source is sent to $0$. An efficient way to do so is to evaluate the effective action\footnote{In principle, one should introduce external sources coupled to the auxiliary fields $c$, $\bar c$ and $h$, which then leads to an effective action $\Gamma_{\bar A}[A,c,\bar c,h]$ whose arguments represent the expectation values of these auxiliary fields in the presence of the sources. In the limit of zero-sources, however, the ghost expectation values vanish which means that one can restrict from the beginning to $\smash{\Gamma_{\bar A}[A,h]\equiv\Gamma_{\bar A}[A,0,0,h]}$. Moreover, it can be shown that the $h$-dependent part of $\Gamma_{\bar A}[A,h]$ is the one already present in the classical action: $\Gamma_{\bar A}[A,h]=\Gamma_{\bar A}[A,0]+\int_x ih^a\bar D_\mu^a(A_\mu^a-\bar A_\mu^a)$. The extremization with respect to $h^a$ leads simply to the gauge condition (\ref{eq:condition}). Thus, in practice, one can assume that $\smash{D_{\mu}[\bar A](A_\mu-\bar A_\mu)=0}$ and, thus, effectively, $\smash{h=0}$.} $\Gamma_{\bar A_c}[A]$ which takes its minimal value when $\smash{A=\langle A\rangle_{\bar A_c}}$.  We have already mentioned that this quantity plays the role of an order parameter for the confinement/deconfinement transition. The argument in terms of the effective action goes as follows.

First, for an arbitrary background $\bar A$, the effective action $\Gamma_{\bar A}[A]$ obeys the identity \cite{Weinberg:1996kr}
\beq
\forall U\in {\cal G}\,, \quad \Gamma_{\bar A^U}[A^U]=\Gamma_{\bar A}[A]\,.\label{eq:bg_sym}
\eeq
In general, $\smash{\bar A^U\neq\bar A}$, and therefore, this identity connects the effective actions in two distinct gauges, characterized by the backgrounds $\bar A$ and $\bar A^U$ respectively. In particular, upon the action of $U$, the absolute minima of $\Gamma_{\bar A}[A]$ are transformed into absolute minima of $\Gamma_{\bar A^U}[A]$, so there is no actual constraint on the location of the minima,\footnote{Such a constraint would make the minima play the role of order parameters for the symmetry at hand.} just a connection between these minima in two different gauges. 

In contrast, within the center-symmetric Landau gauge, corresponding to a choice of background $\bar A_c$, we have
\beq
\Gamma_{\bar A_c}[A^{U_c}]=\Gamma_{\bar A_c}[A]\,,
\eeq
for any transformation $U_c$ complying with Eq.~(\ref{eq:inv}), i.e. that is not just a genuine gauge transformation. In this case, the absolute minima of $\Gamma_{\bar A_c}[A]$ are transformed into absolute minima of the same functional. Within the explicit, Wigner-Weyl realization of the symmetry, the minimum is unique, and, therefore, it needs to be invariant under $U_c$ and thus to correspond to a center-invariant configuration. In contrast, within the broken, Nambu-Goldstone realization of the symmetry, the absolute minima become degenerate, there is no such constraint and the transformation $U_c$ simply transforms the possible minima into one another.

\subsection{Effective potential}
In the case where the background is taken constant, temporal, and Abelian, and because $\langle A_\mu\rangle_{\bar A_c}$ obeys the same properties, it is enough to minimize the effective action within the subspace of configurations (\ref{eq:res}). Up to a trivial space-time volume factor, this amounts to minimizing an effective potential $V_{\bar r_c}(r)$, a function of the $N-1$ real variables $r^j$. 

In practice, it is convenient to first evaluate the potential $V_{\bar r}(r)$ associated to a generic background of the form (\ref{eq:bg}) and only then to restrict to confining backgrounds. At one-loop order, this potential is easily computed since it only requires the evaluation of the determinant of the quadratic part of the gauge-fixed action (\ref{eq:action}) as the fields are varied around configurations that represent the arguments of the effective action. The result of this calculation has already been presented in Ref.~\cite{vanEgmond:2021jyx,Lecture_notes}. The calculation is pretty standard once one introduces the appropriate color basis. For the sake of simplicity, we shall gather the associated details in App.~\ref{app:det} and just recall here the choice of color basis.

In the presence of backgrounds of the form (\ref{eq:bg}), it is convenient to switch from the usual, Cartesian bases $\{t^a\}$ of the color algebra, to Cartan-Weyl bases $\{t^\kappa\}$ such that
\beq
[t^j,t^\kappa]=\kappa^j t^\kappa\,.
\eeq
The labels $\kappa$ are real-valued vectors of $\mathds{R}^{N-1}$ known as adjoint weights. They can be of two types. If they are non-zero, they are called roots and denoted by the first letters of the Greek alphabet: $\alpha$, $\beta$, $\dots$ We stress that roots appear always in pairs, that is if $\alpha$ is a root, then so is $-\alpha$. In the case where the label $\kappa$ vanishes, one needs to add one extra label to indicate that it can correspond to any of the Abelian $t^j$: $\smash{\kappa=0^{(j)}}$ such that $\smash{t^{0^{(j)}}=t^j}$. We refer to $0^{(j)}$ as a zero. In the SU($2$) case for instance, we have one zero and two roots, $\pm1$, while in the SU($3$) case, we have two zeros and six roots $\pm(1,0)$, $\pm (1/2,\sqrt{3}/2)$ and $\pm (1/2,-\sqrt{3}/2)$.

In terms of the adjoint weights $\kappa$, the one-loop potential $V_{\bar r}(r)$ reads, see App.~\ref{app:det},
\begin{eqnarray}
 V_{\bar r}(r) & = & \frac{m^2T^2}{2g^2}(r-\bar r)^2\nonumber\\
& + & \frac{d-2}{2}\sum_\kappa\int_Q^T\ln\big[Q_\kappa^2+m^2\big]\nonumber\\
& + & \frac{1}{2}\sum_\kappa\int_Q^T\ln\left[1+\frac{m^2\bar Q_\kappa^2}{(Q_\kappa\cdot\bar Q_\kappa)^2}\right],\label{eq:final}
\end{eqnarray}
where the last line includes the ghost contribution $-(1/2)\ln\, (Q_\kappa\cdot\bar Q_\kappa)^2$ which rewrites $\smash{-\ln\, |Q_\kappa\cdot\bar Q_\kappa|}$ or, equivalently, $\smash{-{\rm Re}\,\ln\, (Q_\kappa\cdot\bar Q_\kappa)}$. The absolute value or the real part is important since the sign of $Q_\kappa\cdot\bar Q_\kappa$ can vary as one varies $r$ and $\bar r$. In the examples treated below, however, we shall see that, when $\bar r$ is chosen in a given Weyl chamber, the value of $r$ that minimizes $V_{\bar r}(r)$ lies in the same Weyl chamber in such a way that $Q_\kappa\cdot \bar Q_\kappa$ remains positive \cite{vanEgmond:2022nuo,vanEgmond:2023lfu} and one may remove the absolute value/real part, we shall keep it for completeness.

The notation $\int_Q^T$ stands for a Matsubara sum-integral
\beq
\int_Q^T f(Q)\equiv T\sum_{q\in\mathds{Z}}\int\frac{d^{d-1}q}{(2\pi)^{d-1}} f(\omega_q,\vec{q})\,,\label{eq:Mat}
\eeq
with $\smash{Q=(\omega_q,\vec{q})}$ and $\smash{\omega_q=2\pi T q}$ the associated Matsubara frequency.\footnote{We use the same letter $q$ to denote the integer labelling the Matsubara frequencies $\omega_q$ and the associated spatial momentum $\vec{q}$ with norm $q$. This should always be clear from the context.} We mention that, within dimensional regularization, the potential is $\smash{d=4-2\epsilon}$ dimensional, which goes together with the fact that the bare coupling $g^2$ has dimension $2\epsilon$. It is then convenient to multiply the potential by $\Lambda^{2\epsilon}$, with $\Lambda$ an arbitrary scale, so that it becomes $4$-dimensional. The effective way to do so is to redefine the Matsubara sum-integral (\ref{eq:Mat}) to include a factor $\Lambda^{2\epsilon}$ and to view the bare coupling appearing in the tree-level term of Eq.~(\ref{eq:final}) as a dimensionless bare coupling obtained from the dimensional one after applying the rescaling $g^2\to g^2\Lambda^{2\epsilon}$. Of course, physical results should not depend on the scale $\Lambda$.\footnote{This scale is sometimes mistaken with the renormalization scale. This is because, within the minimal subtraction scheme, it coincides with it. But, in general, this scale has to be viewed as a regulating scale, akin to the cut-off in momentum regularizations \cite{Barrios:2021cks}. It differs in general from the renormalization scale $\mu$ that enters the renormalization conditions, and it is actually replaced by the latter in the renormalized expressions, see below.}

As for the other notations in Eq.~(\ref{eq:final}), $(r-\bar r)^2$ designates the square of the vector $r-\bar r$, that is $\sum_j (r_j-\bar r_j)^2$, while $Q_\kappa^2$ and $\bar Q^2_\kappa$ are the squares of the four-momenta
\beq
Q^\kappa_\mu & = & Q_\mu+T(r\cdot\kappa)\delta_{\mu0}\,,\\
\bar Q^\kappa_\mu & = & Q_\mu+T(\bar r\cdot\kappa)\delta_{\mu0}\,,
\eeq
which we refer to as shifted or generalized momenta. Note that the shifts affect only the frequency components and play, therefore, the role of (imaginary) chemical potentials. Aside from the tree-level term in Eq.~(\ref{eq:final}), these shifts are the only source of $r$- and $\bar r$-dependence in the one-loop effective potential (\ref{eq:final}). In fact, only those color labels corresponding to roots carry this dependence. To minimize the potential, we can then restrict the sum over color labels $\kappa$ in Eq.~(\ref{eq:final}) to a sum over the roots $\alpha$. Moreover, since a given root $\alpha$ and the associated root $-\alpha$ contribute the same,\footnote{This is easily shown by making the change of variables $\smash{\omega_q\to-\omega_q}$ under the Matsubara sum.} one can just sum over half the roots and omit the factors of $1/2$ in the last two terms of Eq.~(\ref{eq:final}).

\subsection{Symmetries}\label{sec:sym}

Let us now review some of the symmetries of the effective potential $V_{\bar r}(r)$.

\subsubsection{Gauge transformations}
To each root $\alpha$, one can associate two particular transformations within ${\cal G}_0$ with the additional property that they leave the particular form of the background (\ref{eq:bg}) invariant \cite{Reinosa:2020mnx,vanEgmond:2023lfu}, with possibly a different value of $\bar r$. The first of these transformations is a color rotation, known as Weyl transformation, that acts on (\ref{eq:bg}) as a reflection with respect to a hyperplane orthogonal to $\alpha$:
\beq
\bar r\to \bar r-2\frac{\bar r\cdot\alpha}{\alpha^2}\alpha\,.\label{eq:Weyl}
\eeq
The other transformation corresponds to a translation by a vector $4\pi\alpha$:
\beq
\bar r\to \bar r+4\pi\alpha\,.\label{eq:winding}
\eeq
As a consequence of the identity (\ref{eq:prop}), the effective potential should then obey
\beq
V_{\bar r}(r) & = & V_{\bar r-2\frac{\bar r\cdot\alpha}{\alpha^2}\alpha}\left(r-2\frac{r\cdot\alpha}{\alpha^2}\alpha\right)\nonumber\\
& = & V_{\bar r+4\pi\alpha}(r+4\pi\alpha)\,.\label{eq:sym2}
\eeq
We can readily check these properties on the one-loop expression (\ref{eq:final}) given above. For instance, under the transformation (\ref{eq:winding}), we have
\beq
Q^\kappa_\mu\to Q^\kappa_\mu+T(4\pi\alpha\cdot\kappa)\delta_{\mu0}\,,\label{eq:shift1}
\eeq
and similarly for $\bar Q_\kappa$. Now, it can be shown that the only possible values taken by $\alpha\cdot\kappa$ are $-1,-1/2,0,+1/2,+1$, see for instance Ref.~\cite{vanEgmond:2023lfu}. Therefore, for each $\kappa$, the frequency shift in Eq.~(\ref{eq:shift1}) is either zero or corresponds exactly to one or two Matsubara frequencies. This means that, in each term of Eq.~(\ref{eq:final}), the shift can be reabsorbed through a change of variables in the corresponding Matsubara sum, which leads then to the second identity in Eq.~(\ref{eq:sym2}). In the SU($2$) case for instance, this identity writes $\smash{V_{\bar r}(r)=V_{\bar r+4\pi}(r+4\pi)}$.

Similarly, under the transformation (\ref{eq:Weyl}), we have
\beq
Q^\kappa_\mu\to Q^\kappa_\mu-2T\frac{\bar r\cdot\alpha\,\alpha\cdot\kappa}{\alpha^2}\delta_{\mu0}=Q^{\kappa-2\frac{\kappa\cdot\alpha}{\alpha^2}\alpha}_\mu\,.
\eeq
It can also be shown that $\kappa-2(\kappa\cdot\alpha/\alpha^2)\alpha$ is either $0$ (when $\kappa$ is a zero) or spans all possible roots as $\kappa$ spans the roots, see \cite{vanEgmond:2023lfu}. This means that the terms in Eq.~(\ref{eq:final}) are simply reshuffled into one another by the considered transformation, which leads to the first identity in Eq.~(\ref{eq:sym2}). In the SU($2$) case, this identity becomes $\smash{V_{\bar r}(r)=V_{-\bar r}(-r)}$.

\subsubsection{Center transformations}
So far, we considered transformations within ${\cal G}_0$, that is transformation corresponding to $\smash{k=0}$ in Eq.~(\ref{eq:k}). There are also transformations with $\smash{k\neq 0}$. They are all generated by 
\beq
\bar r\to \bar r+4\pi\rho\,,\label{eq:center}
\eeq
corresponding to $\smash{k=1}$. Here, the $\rho$'s denote the defining weights of SU(N), obtained after diagonalizing the defining action of the algebra, $\smash{t^j|\rho\rangle=\rho^j|\rho\rangle}$. We should again have
\beq
V_{\bar r}(r) & = & V_{\bar r+4\pi\rho}(r+4\pi\rho)\,,\label{eq:sym_rho}
\eeq
which is readily checked using the same argument as above, combined with the fact that the only possible values for $\kappa\cdot\rho$ are $-1/2,0,+1/2$. In the SU($2$) case in particular, we should have $\smash{V_{\bar r}(r)=V_{\bar r+2\pi}(r+2\pi)}$.

\subsubsection{Charge conjugation}
Charge conjugation acts on $(r,\bar r)$ as $(-r,-\bar r)$. It follows that
\beq
V_{\bar r}(r)=V_{-\bar r}(-r)\,.
\eeq
This can be explicitly verified on Eq.~(\ref{eq:final}) since the tree-level term is quadratic in $r-\bar r$ and the one-loop contribution associated to a mode $\kappa$ depends on $r$ and $\bar r$ via the combinations $\kappa\cdot r$ and $\kappa\cdot\bar r$ in such a way that the transformation $\smash{(r,\bar r)\to (-r,-\bar r)}$ relates the contributions from the modes $\kappa$ and $-\kappa$. It is even simpler than that because, as we have seen, the contributions from the modes $\kappa$ and $-\kappa$ are identical and, thus, each mode contribution is invariant under $\smash{(r,\bar r)\to (-r,-\bar r)}$.

\subsubsection{Weyl chambers}
The transformations (\ref{eq:Weyl})-(\ref{eq:winding}) are also intimately related to the Weyl chambers alluded to above and which, among other things, allow one to identify center-invariant configurations as defined in Eq.~(\ref{eq:inv}). More precisely, by combining these transformations, one generates reflections orthogonal to a given root $\alpha$ and displaced with respect to the origin by any multiple of $2\pi$ times that root. The Weyl chambers appear as the regions delimited by the hyperplanes associated with all these reflections for all the roots of the algebra \cite{Reinosa:2020mnx,vanEgmond:2023lfu}.

Now, under a center transformation (\ref{eq:center}), a given Weyl chamber is transformed into a different one. However, upon using the gauge transformations that connect the Weyl chambers into one another, one can bring the transformed Weyl chamber back on top of the original one. In doing so, one obtains a transformation of a Weyl chamber into itself whose fixed points are center invariant configurations in the sense of Eq.~(\ref{eq:inv}). In the SU($2$) case for instance, the Weyl chambers are the intervals $[2\pi n,2\pi(n+1)]$ and the action of a center transformation with $\smash{k=1}$ on a given Weyl chamber is a reflection about its center $2\pi(n+1/2)$.

Turning back to the effective potential $V_{\bar r}(r)$, we have seen that, when choosing $\smash{\bar r=\bar r_c}$ center-invariant and as long as center symmetry is not spontaneously broken, the minimum of $V_{\bar r_c}(r)$ needs to be center-invariant as well. More precisely it needs to be invariant under the same transformation $U_c$ that leaves $\bar r_c$ invariant. Now, since this transformation has usually only one fixed-point \cite{vanEgmond:2023lfu}, it follows that, as long as center symmetry is not broken, the minimum of $V_{\bar r_c}(r)$ needs to be $\smash{r=\bar r_c}$.

Similar considerations, allow one to identify charge-conjugation invariant configurations (modulo gauge transformations). In the SU($2$) case, it is found that any configuration is invariant in this sense under charge conjugation. In the SU($3$) case, in contrast, not all configurations are compatible with charge-conjugation. Among the configurations of the form (\ref{eq:bg}), one example is provided by those with $\smash{\bar r^8=0}$ which are invariant under $\smash{\bar r^8\to -\bar r^8}$. For backgrounds of this form, and because charge conjugation is not expected to break spontaneously, the one-point function needs also to be invariant under $\smash{r^8\to -r^8}$ and thus $\smash{r^8=0}$. We shall use this remark below to simplify the discussion in the SU($3$) case.

\section{Renormalization}\label{sec:renormalization}
Before considering the practical evaluation of the potential, we analyze its divergences and their renormalization.

\subsection{General considerations}
Let us start by emphasizing that the analysis of divergences in the presence of a background slightly differs from the corresponding analysis in the absence of a background. Yet, in the case of backgrounds of the form (\ref{eq:bg}), the two can be easily connected. 

One first notices that, in deriving Eq.~(\ref{eq:bg_sym}), one usually assumes that the gauge fields are periodic, and, consequently, the considered transformations $U$ need to belong to ${\cal G}$ for the periodicity to be preserved. It is possible, however, to consider transformations with other boundary conditions, the only change in Eq.~(\ref{eq:bg_sym}) being that the transformed gauge fields are periodic up to the considered transformation:\footnote{Not only are the background $\bar A^U$ and the argument $A^U$ of the effective action periodic up to the transformation $U$, but also the gauge-fields over which one integrates to evaluate the various contributions to the effective action, hence the notation $\Gamma^U$.}
\beq
\Gamma_{\bar A}[A]=\Gamma^U_{\bar A^U}[A^U]\,.\label{eq:gen}
\eeq
In particular, for backgrounds of the form (\ref{eq:bg}), one can take the transformation $\bar U\equiv e^{-i(\tau/\beta) \bar r^j t^j}$ such that
\beq
\bar A_\mu^{\bar U} & = & e^{-i\frac{\tau}{\beta} \bar r^j t^j}\frac{T}{g}\delta_{\mu0} \bar r^j t^j e^{i\frac{\tau}{\beta} \bar r^j t^j}\nonumber\\
& + & \frac{i}{g}e^{-i\frac{\tau}{\beta} \bar r^j t^j}\partial_\mu e^{i\frac{\tau}{\beta} \bar r^j t^j}=0\,.
\eeq
Then, from Eq.~(\ref{eq:gen}), we find
\beq
\Gamma_{\bar A}[A]=\Gamma^{\bar U}_{0}[A^{\bar U}]\,.\label{eq:gen2}
\eeq
We can also rewrite $A^{\bar U}$ as 
\beq
A^{\bar U} & = & (\bar A+A-\bar A)^{\bar U}\nonumber\\
& = & \bar A^{\bar U}+\bar U(A-\bar A){\bar U}^\dagger\nonumber\\
& = & \bar Ua{\bar U}^\dagger\,,
\eeq
with $\smash{a\equiv A-\bar A}$. We have thus shown that
\beq
\Gamma_{\bar A}[A]=\Gamma^{\bar U}_{0}[\bar Ua {\bar U}^\dagger]\,.\label{eq:gen3}
\eeq
Since the boundary conditions should not affect the UV divergences, it follows, as announced, that the analysis of the UV divergences of $\Gamma_{\bar A}$ boils down to that of the UV divergences of $\Gamma_{\bar A=0}$. In particular, the elimination of UV divergences requires the multiplicative renormalization of $a$:
\beq
a\to \sqrt{Z_a} a\,,
\eeq
rather than $A$, see below for further remarks.

Also, since $\bar U$ plays a spectator role with regard to the UV divergences in Eq.~(\ref{eq:gen3}), it remains finite upon renormalization, and because, it depends only on the combination $\smash{\bar r\propto g\bar A}$, we deduce that the renormalization of the coupling and the renormalization of the background are intimately related. More precisely, the corresponding renormalization factors $Z_{g^2}$ and $Z_{\bar A}$ are such that their product $Z_{g^2}Z_{\bar A}$ is finite. As a consequence, one can choose to work within renormalization schemes where
\beq
Z_{\bar A}=Z_{g^2}^{-1}\,.\label{eq:renorm}
\eeq 
In the following, we restrict ourselves to using these schemes, and hence the combination $g\bar A$ (and therefore $\bar r$) does not renormalize. We stress that, in contrast, the renormalization of $a$ cannot be entirely encoded in that of $g$ since the second derivative of the effective action with respect to $a$ is nothing but the inverse propagator, and it is well known that the latter renormalizes differently than the coupling. In other words, the combination $g\,a$ (and therefore $r-\bar r$) does get renormalized.

Still, one could wonder why it is not possible to assume that $\bar A$ and $A$ renormalize both multiplicatively, that is $\smash{\bar A\to \sqrt{Z_{\bar A}}\bar A}$ and $\smash{A\to \sqrt{Z_A} A}$, with $\smash{Z_{\bar A}\neq Z_A}$. This choice is too na\"\i ve, however, because this implies that the effective action for the rescaled fields would obey (\ref{eq:bg_sym}) with
\beq
\bar A_\mu(x) & \!\to\! &  U(x)\bar A_\mu(x)U^\dagger(x)\!+
\!\frac{i}{g}U(x)\partial_\mu U^\dagger(x)\,,\\
A_\mu(x) & \!\to\! &  U(x)A_\mu(x)U^\dagger(x)\!+\!\sqrt{\frac{Z_{\bar A}}{Z_A}}\frac{i}{g}U(x)\partial_\mu U^\dagger(x)\,,\nonumber\\
\eeq
where we have used Eq.~(\ref{eq:renorm}) and $g$ is now the renormalized coupling. This is problematic, however, because the presence of $Z_{\bar A}/Z_A$ makes the transformation of $A$ ill-defined. On the other hand, assuming that the fields that renormalize multiplicatively are not $\bar A$ and $A$ but rather $\bar A$ and $\smash{a\equiv A-\bar A}$:
\beq
\bar A\to \sqrt{Z_{\bar A}}\bar A\,, \quad a\to \sqrt{Z_a} a\,,\label{eq:resc}
\eeq
and defining $\smash{\hat\Gamma_{\bar A}[a]\equiv\Gamma_{\bar A}[A]}$, one finds
\beq
\hat\Gamma_{\bar A^U}[Ua U^\dagger]=\hat\Gamma_{\bar A}[a]\,,
\eeq 
which remains unchanged upon rescaling $\bar A$ and $a$ according to Eq.~(\ref{eq:resc}).

\subsection{Divergences}

The above considerations lead naturally to the conclusion that the effective action should be more conveniently seen as a functional of $\bar A$ and $a$. The UV divergences are entirely contained in the first terms of the Taylor expansion of $\hat\Gamma_{\bar A}[a]$ around $\smash{a=0}$ and relate to the zero-, two-, three and four-gluon vertex functions. In the case of the one-loop potential $V_{\bar r}(r)$, since the field and the background are taken in the commuting part of the algebra, the divergences associated to three and four-gluon functions should not be present since there would be no tree-level term to absorb them. To capture all the UV divergences, it is then enough to determine the Taylor expansion up to second order. 

To do so, it is convenient to first express the potential in terms of renormalized variables defined by the re-scalings
\begin{equation}
m^2\to Z_{m^2} m^2, \,\, g^2\to Z_{g^2} g^2\,, \,\, \bar r\to\bar r, \,\, \Delta r \to Z_{g^2}^{1/2}Z_a^{1/2}\Delta r\,,
\end{equation}
where we have used that $\bar r\propto g\bar A$ does not renormalize, as well as $\smash{\Delta r\propto g(A-\bar A)}$. Then, the Taylor expansion of the effective potential around the background to second order in the renormalized $\Delta r$ reads\footnote{To this order of accuracy in perturbation theory, renormalization factors appear only in the tree-level contributions, when any. They are set to $1$ within the one-loop contributions.}
\beq
[V_{\bar r}(r)]_2 & \equiv & V_{\bar r}(\bar r)+\sum_j \left.\frac{\partial V_{\bar r}(r)}{\partial r^j}\right|_{r=\bar r}\Delta r^j\nonumber\\
& + & \frac{1}{2} \sum_{j,k} \left.\frac{\partial^2 V_{\bar r}(r)}{\partial r^j\partial r^k}\right|_{r=\bar r}\Delta r^j \Delta r^k\,,\label{eq:taylor}
\eeq
with
\beq
V_{\bar r}(\bar r)=\sum_\kappa\left[\frac{d-1}{2}\int_Q^T\ln\big[\bar Q_\kappa^2+m^2\big]-\frac{1}{2}\int_Q^T\ln \bar Q_\kappa^2\right],\label{eq:V0}\nonumber\\
\eeq
\beq
\left.\frac{\partial V_{\bar r}(r)}{\partial r^j}\right|_{r=\bar r}\!\!\!\!=T\sum_\kappa \kappa^j \left[(d-1)\int_Q^T\!\frac{\bar\omega_q^\kappa}{\bar Q_\kappa^2+m^2}-\!\int_Q^T\!\frac{\bar\omega_q^\kappa}{\bar Q_\kappa^2}\right],\label{eq:V1}\nonumber\\
\eeq
and
\beq
& & \left.\frac{\partial^2 V_{\bar r}}{\partial r^j\partial r^k}\right|_{r=\bar r}\nonumber\\
& & \hspace{0.5cm}=\,T^2\Bigg[Z_a Z_{m^2}\frac{m^2}{g^2}\delta^{jk}+\sum_\kappa \kappa^j\kappa^k\nonumber\\
& & \hspace{0.7cm}\times\,\left( \int^T_Q \frac{(\bar\omega_q^\kappa)^2}{\bar Q_\kappa^4}-2(d-1)\int^T_Q\frac{(\bar\omega_q^\kappa)^2}{(\bar Q_\kappa^2+m^2)^2}\right.\nonumber\\
& & \hspace{0.9cm}+\,\left.(d-2)\int^T_Q\frac{1}{\bar Q_\kappa^2+m^2}+\int^T_Q\frac{(\bar\omega_q^\kappa)^2}{\bar Q^2_\kappa(\bar Q^2_\kappa+m^2)}\right)\Bigg].\label{eq:V2}\nonumber\\
\eeq
The above expressions involve one-loop Matsubara sum-integrals which can here all be split into a vacuum contribution, defined as the $\smash{T\to 0}$ limit of the corresponding expressions with $T\bar r$ kept fixed, and a thermal contribution, see Appendix \ref{app:sums} for more details.\footnote{There, we also discuss one example where the splitting is not possible.} We can denote this splitting formally as
\beq
\int^T_Q\,f(Q)=\int_Q\,f(Q)+\int_Q^{\rm th}\,f(Q)\,,\label{eq:splitting}
\eeq 
where the last term should actually be seen as a $q$-integral (obtained after performing the Matsubara sums) but it is convenient to denote it formally as a $Q$-sum-integral, see below. The UV divergences are entirely contained within the vacuum contributions, so we can ignore the thermal contributions for now. We shall evaluate them later.

The vacuum contributions are simply obtained by replacing the discrete Matsubara summations by continuous frequency integrals. In this case, the color-dependent frequency shifts can be absorbed via a change of variables, and, therefore, the sums become independent of $\bar r$. For instance, the vacuum contribution to (\ref{eq:V0}) is
\beq
\frac{d-1}{2}(N^2-1)\int_Q\ln\big[Q^2+m^2\big]\,,
\eeq
where we used $\smash{\sum_\kappa 1=N^2-1}$ and $\smash{\int_Q \ln Q^2=0}$.
In principle, the corresponding divergence needs to be absorbed in a shift of the potential. However, because we are only interested in extremizing with respect to $r$, we can ignore this contribution. As for the vacuum contribution in Eq.~(\ref{eq:V1}), it is seen to vanish trivially. This comes either from $\smash{\int_Q \omega_q/(Q^2+m^2)=0}$ or from $\smash{\sum_\kappa \kappa^j=0}$. There is thus no divergence within (\ref{eq:V1}), in line with the fact that there is no tree-level term to absorb it. 

We are then left with the vacuum contribution in Eq.~(\ref{eq:V2}). It can be given a simple form by using $\int_Q \omega^2_q/Q^4=(1/d)\int_Q 1/Q^2=0$ as well as
\beq
\int_Q \frac{\omega_q^2}{(Q^2+m^2)^2} & = & \frac{1}{d}\int_Q \frac{Q^2}{(Q^2+m^2)^2}\nonumber\\
& = & \frac{1}{d}\int_Q \left[\frac{1}{Q^2+m^2}-\frac{m^2}{(Q^2+m^2)^2}\right]\nonumber\\
& = & \frac{1}{d}\int_Q \frac{1}{Q^2+m^2}\left[1-\frac{\Gamma(2-d/2)}{\Gamma(1-d/2)}\right]\nonumber\\
& = & \frac{1}{2} \int_Q \frac{1}{Q^2+m^2}\,,\label{eq:44}
\eeq
and
\beq
\int_Q\frac{\omega_q^2}{Q^2(Q^2+m^2)} & = & \frac{1}{d}\int_Q\frac{1}{Q^2+m^2}\,.\label{eq:45}
\eeq
Altogether, the vacuum contribution to the square bracket in the RHS of Eq.~(\ref{eq:V2}) reads $(M^2_{T=0}/g^2)\delta_{jk}$, with
\beq
M^2_{T=0} & = & Z_a Z_{m^2}m^2-\frac{d-1}{d}g^2N\int_Q\frac{1}{{Q}^2+m^2}\nonumber\\
 & = & Z_a Z_{m^2}m^2+\frac{3-2\epsilon}{4-2\epsilon}\frac{g^2N}{16\pi^2}m^2\left[\frac{1}{\epsilon}+\ln\frac{\bar\Lambda^2}{m^2}+1\right]\nonumber\\
& = & \left[Z_a Z_{m^2}+\frac{3g^2N}{64\pi^2}\left(\frac{1}{\epsilon}+\ln\frac{\bar\Lambda^2}{m^2}+\frac{5}{6}\right)\right]m^2,\label{eq:Mvac}
\eeq
where $\smash{\bar\Lambda^2\equiv 4\pi\Lambda^2 e^{-\gamma}}$ and we have used $\smash{\sum_\kappa \kappa^j\kappa^k=N\delta_{jk}}$. The notation $M^2_{T=0}$ is not innocent. This is because, as it can be easily argued, within the gauge $\smash{\bar r=\bar r_c}$ and as long as the system is in the symmetric phase for which $\smash{r=\bar r_c}$, the quantity $(g^2/T^2)\partial^2 V_{\bar r}(r)/\partial r^j\partial r^k|_{r=\bar r}$, is nothing but the zero-temperature, zero-momentum mass as obtained from the inverse gluon propagator in this limit. 

As a check of Eq.~(\ref{eq:Mvac}), we notice that, writing the renormalization factors as $\smash{Z=1+\delta Z}$, we deduce that
\beq
\delta Z_a+\delta Z_{m^2}=-\frac{3g^2N}{64\pi^2}\frac{1}{\epsilon}+\cdots
\eeq
This is in agreement with the known divergent contributions for $Z_a$ and $Z_{m^2}$ obtained from the vacuum propagator at one-loop order \cite{Tissier:2011ey,Gracey:2001ma}. Thus the divergent contribution to the potential is correctly renormalized at one-loop order.  In the next section, we explain how to evaluate the corresponding finite contribution. Of course, part of this contribution has to do with fixing the finite parts of the renormalization factors. This we do in Sec.~\ref{sec:schemes} where we review various possible renormalization schemes together with the associated renormalization group flow.

\section{Evaluation of the potential}\label{sec:evaluation}
Let us now detail the evaluation of the potential. Since we have already gone through the extraction of the UV divergences, it is actually simpler to organize the calculation as
\beq
V_{\bar r}(r)=[V_{\bar r}(r)]_2+\delta V_{\bar r}(r)\,,
\eeq
with
\beq
\delta V_{\bar r}(r)\equiv V_{\bar r}(r)-[V_{\bar r}(r)]_2\,.
\eeq
The divergences are entirely contained within the vacuum contribution to $[V_{\bar r}(r)]_2$ which we have computed in the previous section, and which is given by 
\beq
V^{T=0}_{\bar r}(\bar r)+\frac{T^2}{2g^2}M^2_{T=0}(\Delta r)^2\,,
\eeq
where we recall that the first term can be ignored for it does not depend on $r$. We are thus left with the determination of both the thermal contribution to $[V_{\bar r}(r)]_2$ and the UV-finite difference $\delta V_{\bar r}(r)$.

\subsection{Thermal contribution to $[V_{\bar r}(r)]_2$}
As before, we can ignore the thermal contribution to $V_{\bar r}(\bar r)$ since it does not depend on $r$. This function, however, will re-enter our discussion below for it is at the basis of a popular approach to which we shall compare our results. 

The first derivative of the potential appearing in Eq.~(\ref{eq:taylor}) is purely thermal as we have seen. Performing the Matsubara sums, see App.~\ref{app:sums}, one finds
\beq
\left.\frac{\partial V_{\bar r}(r)}{\partial r^j}\right|_{r=\bar r}=\frac{T}{2\pi^2}\sum_\kappa \kappa^j \int_0^\infty \!\!\!dq\,q^2\,{\rm Im}\,\big[3n_{\bar\varepsilon^\kappa_q}-n_{\bar q^\kappa}\big]\,,\label{eq:V11}\nonumber\\
\eeq
where we have set $\smash{d=4}$ (since this contribution is UV finite) and we have defined the Bose-Einstein distribution $\smash{n_\varepsilon\equiv 1/(e^{\beta\varepsilon}-1)}$ as well as $\smash{\bar\varepsilon_q^\kappa\equiv\varepsilon_q-iT\bar r\cdot\kappa}$ and $\smash{\varepsilon_q\equiv\sqrt{q^2+m^2}}$. 

The second derivative of the potential appearing in Eq.~(\ref{eq:taylor}) rewrites as $M^2_{T,jk}T^2/g^2$, where $M^2_{T,jk}$ is the zero-momentum mass whose vacuum contribution was determined above. To evaluate the thermal contribution, we use\footnote{Here it is important to stress that the notation $\int_Q^{\rm th}$ introduced in Eq.~(\ref{eq:splitting}) is actually a formal proxy for a $q$-integral rather than an actual $Q$-sum-integral. However, as long as the considered manipulations do not involve frequencies, we can perform these manipulations at the level of the formal notation $\int_Q^T$.}
\beq
& & \int_Q^{\rm th}\frac{(\bar\omega_q^\kappa)^2}{\bar Q^2_\kappa(\bar Q^2_\kappa+m^2)}\nonumber\\
& & \hspace{1.0cm}\,=\, \int_Q^{\rm th}\frac{1}{m^2}\left[\frac{(\bar\omega_q^\kappa)^2}{\bar Q^2_\kappa}-\frac{(\bar\omega_q^\kappa)^2}{\bar Q^2_\kappa+m^2}\right]\nonumber\\
& & \hspace{1.0cm}\,=\,\int_Q^{\rm th}\frac{1}{m^2}\left[\frac{q^2+m^2}{\bar Q^2_\kappa+m^2}-\frac{q^2}{\bar Q^2_\kappa}\right],\label{eq:52}
\eeq
and
\beq
& & \int_Q^{\rm th}\frac{(\bar\omega_q^\kappa)^2}{({\bar Q}_\kappa^2+m^2)^2}\nonumber\\
& & \hspace{1.0cm}\,=\,\int_Q^{\rm th}\left[\frac{1}{{\bar Q}_\kappa^2+m^2}-\frac{q^2+m^2}{({\bar Q}_\kappa^2+m^2)^2}\right]\nonumber\\
& & \hspace{1.0cm}\,=\,\int_Q^{\rm th}\left[\frac{1}{{\bar Q}_\kappa^2+m^2}+\frac{q^2+m^2}{2q}\frac{d}{dq}\frac{1}{{\bar Q}_\kappa^2+m^2}\right]\nonumber\\
& & \hspace{1.0cm}\,=\,\int_Q^{\rm th}\left[\frac{1}{{\bar Q}_\kappa^2+m^2}-\frac{1}{2}\left(3+\frac{m^2}{q^2}\right)\frac{1}{{\bar Q}_\kappa^2+m^2}\right]\nonumber\\
& & \hspace{1.0cm}\,=\,-\frac{1}{2}\int_Q^{\rm th}\left(1+\frac{m^2}{q^2}\right)\frac{1}{{\bar Q}_\kappa^2+m^2}\,.\label{eq:53}
\eeq
Because we focused on the thermal contributions only,   we could set $\smash{d=4}$ from the start and we could safely neglect the boundary terms in the integration by parts used in the last steps of (\ref{eq:53}).\footnote{With some effort, it is possible to show that, in dimensional regularization, the same manipulations hold for the vacuum contributions.}  Putting all the pieces together, one finds the thermal contribution
\beq
\delta M^2_{{\rm th},jk} & \,=\, & g^2\sum_\kappa\kappa^j\kappa^k\left[\int_Q^T\left(3\frac{m^2}{q^2}+6+\frac{q^2}{m^2}\right)\frac{1}{\bar Q^2_\kappa+m^2}\right.\nonumber\\
& & \hspace{2.5cm}\left.-\,\int_Q^T\left(\frac{1}{2}+\frac{q^2}{m^2}\right)\frac{1}{\bar Q^2_\kappa}\right],\label{eq:MT2}
\eeq
to which we should of course add the expression for $M^2_{T=0}\delta_{jk}$ obtained above, see Eq.~(\ref{eq:Mvac}). In this form, the Matsubara sums can be simply performed, see App.~\ref{app:sums}, and we find
\beq
M^2_{T,jk} & = & \left[Z_a Z_{m^2}+\frac{3g^2N}{64\pi^2}\left(\frac{1}{\epsilon}+\ln\frac{\bar\Lambda^2}{m^2}+\frac{5}{6}\right)\right]m^2\delta_{jk}\nonumber\\
& + & \frac{g^2}{2\pi^2}\sum_\kappa\kappa^j\kappa^k\int_0^\infty dq\,q^2\,{\rm Re}\nonumber\\
& & \times\,\left[\left(3\frac{m^2}{q^2}+6+\frac{q^2}{m^2}\right)\frac{n_{\bar\varepsilon_q^\kappa}}{\varepsilon_q}-\left(\frac{1}{2}+\frac{q^2}{m^2}\right)\frac{n_{\bar q^\kappa}}{q}\right].\label{eq:curv3}\nonumber\\
\eeq
This curvature mass matrix is intimately related to the confinement/deconfinement transition. In the SU($2$) case for instance, its vanishing (in the gauge $\smash{\bar r=\pi}$) allows one to extract the transition temperature, whereas in the SU($3$) case, it gives access to the higher spinodal temperature, see Sec.~\ref{sec:results} for more details. For completeness, we mention that the curvature can be put in the following form
\beq
& & M^2_{T,jk}=\sum_\kappa\kappa^j\kappa^k\nonumber\\
& & \times\left[\left(\frac{Z_a Z_{m^2}}{N}+\frac{3g^2}{64\pi^2}\left(\frac{1}{\epsilon}+\ln\frac{\bar\Lambda^2}{m^2}+\frac{5}{6}\right)\right)m^2\right.\nonumber\\
& & \hspace{0.5cm}+\,\frac{g^2}{2\pi^2}\int_0^\infty dq\,q^2\,{\rm Re}\left(3\frac{m^2}{q^2}+6+\frac{q^2}{m^2}\right)\frac{n_{\bar\varepsilon_q^\kappa}}{\varepsilon_q}\nonumber\\
& & \left.\hspace{0.5cm}-\,g^2T^2\left(\frac{1}{4}B_2\left(\left\{\frac{\kappa\cdot\bar r}{2\pi}\right\}\right)-\pi^2B_4\left(\left\{\frac{\kappa\cdot\bar r}{2\pi}\right\}\right)\frac{T^2}{m^2}\right)\right],\label{eq:curv4}\nonumber\\
\eeq
where
\beq
B_2(x) & = & x^2-x+\frac{1}{6}
\eeq
and
\beq
B_4(x) & = & x^4-2x^3+x^2-\frac{1}{30}
\eeq
are the Bernouilli polynomials of degree $2$ and $4$ respectively and $\{x\}$ is the real number between $0$ and $1$ obtained from $x$ by adding the appropriate integer,\footnote{This definition is ambiguous if $x$ is already an integer. However, $\kappa\cdot \bar r/(2\pi)$ will not be chosen to be an integer, and $\kappa\cdot r/(2\pi)$ will turn out not to be an integer either.} see App.~\ref{app:sums}.

\subsection{The UV finite $\delta V_{\bar r}(r)$}
By construction, $\delta V_{\bar r}(r)$ is UV finite. In the present case, it is a pure one-loop contribution which we write for convenience as
\beq
\delta V_{\bar r}(r) & = & \sum_\kappa\int_Q^T \!\Bigg[L(\Delta r)-L(0)-\Delta r^j L'_j(0)\nonumber\\
& & \hspace{2.0cm}-\,\frac{\Delta r^j\Delta r^k}{2}L''_{jk}(0)\Bigg],
\eeq
with
\beq
L(\Delta r)=\frac{1}{2}\ln\left[1+\frac{m^2\bar Q_\kappa^2}{(Q_\kappa\cdot\bar Q_\kappa)^2}\right]+\frac{d-2}{2}\ln\Big[Q^2_\kappa+m^2\Big]\,,\nonumber\\
\eeq
as well as
\beq
L(0) & = & \frac{d-1}{2}\ln\Big[\bar Q^2_\kappa+m^2\Big]-\frac{1}{2}\ln \bar Q^2_\kappa\,,\\
L'_j(0) & = & T\kappa^j \left[(d-1)\frac{\bar\omega_q^\kappa}{\bar Q_\kappa^2+m^2}-\frac{\bar\omega_q^\kappa}{\bar Q_\kappa^2}\right],\\
L''_{jk}(0) & = & T^2\kappa^j\kappa^k\left[ \frac{(\bar\omega_q^\kappa)^2}{\bar Q_\kappa^4}-2(d-1)\frac{(\bar\omega_q^\kappa)^2}{(\bar Q_\kappa^2+m^2)^2}\right.\nonumber\\
& & \hspace{1.2cm}+\left.\frac{d-2}{\bar Q_\kappa^2+m^2}+\frac{(\bar\omega_q^\kappa)^2}{\bar Q^2_\kappa(\bar Q^2_\kappa+m^2)}\right]\!.\nonumber\\
\eeq
In principle, we could apply the strategy followed in the previous subsection, based on performing the Matsubara sums analytically and the resulting momentum integrals numerically. However, some of the Matsubara sums are cumbersome to evaluate using contour integration techniques due to the presence of quartic polynomials in the Matsubara frequencies with no obvious roots. For this reason, we adopt a different strategy based, instead, on performing the momentum integrals analytically and the resulting Matsubara sums numerically.

The first step is to rewrite the momentum integrals as vacuum $\smash{D\equiv d-1}$ integrals. To this purpose, we notice that
\beq
\bar Q^2_\kappa & = & q^2+\bar M^2_{0,\kappa}\,,\\
Q_\kappa\cdot \bar Q_\kappa & = & q^2+M^2_{0,\kappa}\,,\\
\bar Q^2_\kappa+m^2 & = & q^2+\bar M^2_{\kappa}\,,\\
Q^2_\kappa+m^2 & = & q^2+M^2_{\kappa}\,,
\eeq
where we have defined the frequency-dependent masses
\beq
\bar M^2_{0,\kappa} & \equiv & (\bar\omega_q^\kappa)^2\,,\\
M^2_{0,\kappa} & \equiv & \omega_q^\kappa\bar\omega_q^\kappa\,,\\
\bar M^2_{\kappa} & \equiv & (\bar\omega_q^\kappa)^2+m^2\,,\\
M^2_{\kappa} & \equiv & (\omega_q^\kappa)^2+m^2\,.
\eeq
Similarly,
\beq
& & (Q_\kappa\cdot \bar Q_\kappa)^2+m^2\bar Q^2_\kappa\nonumber\\
& & \hspace{0.5cm}=\,q^4+q^2(m^2+2\omega^\kappa_q\bar\omega^\kappa_q)+(\bar\omega^\kappa_q)^2(m^2+(\omega^\kappa_q)^2)\,,\nonumber\\
& & \hspace{0.5cm}=\,(q^2+M^2_{+,\kappa})(q^2+M^2_{-,\kappa})\,,
\eeq
with
\beq
M^2_{\pm,\kappa}\equiv \omega^\kappa_q\bar\omega^\kappa_q+\frac{m^2}{2}\pm\frac{m^2}{2}\sqrt{1+4\frac{\bar\omega^\kappa_q(\omega^\kappa_q-\bar \omega^\kappa_q)}{m^2}}\,.\nonumber\\
\eeq
Then,
\beq
L(\Delta r) & = & \frac{1}{2}\ln\frac{(q^2+M^2_{+,\kappa})(q^2+M^2_{-,\kappa})}{(q^2+M^2_{0,\kappa})^2}\nonumber\\
& & +\,\frac{D-1}{2}\ln(q^2+M^2_{\kappa})\,,
\eeq
and
\beq
L(0) & = & \frac{D}{2}\ln(q^2+\bar M^2_{\kappa})-\frac{1}{2}\ln (q^2+\bar M^2_{0,\kappa})\,,\\
L'_j(0) & = & T\kappa^j \left[D\frac{\bar\omega_q^\kappa}{q^2+\bar M^2_{\kappa}}-\frac{\bar\omega_q^\kappa}{q^2+\bar M^2_{0,\kappa}}\right],\\
L''_{jk}(0) & = & T^2\kappa^j\kappa^k\left[ \frac{(\bar\omega_q^\kappa)^2}{(q^2+\bar M^2_{0,\kappa})^2}-2D\frac{(\bar\omega_q^\kappa)^2}{(q^2+\bar M^2_{\kappa})^2}\right.\nonumber\\
& & \hspace{0.9cm}+\,\left.\frac{D-1}{q^2+\bar M^2_{\kappa}}+\frac{(\bar\omega_q^\kappa)^2}{(q^2+\bar M^2_{0,\kappa})(q^2+\bar M^2_{\kappa})}\right]\!.\nonumber\\
\eeq
Putting all the pieces together, we arrive at
\begin{widetext}
\beq
& & \delta V_{\bar r}(r)=T\sum_\kappa \sum_{q\in\mathds{Z}}\int\frac{d^Dq}{(2\pi)^D}\left[\frac{1}{2}\ln \frac{(q^2+M^2_{+,\kappa})(q^2+M^2_{-,\kappa})(q^2+\bar M^2_{0,\kappa})(q^2+M^2_{\kappa})^{D-1}}{(q^2+M^2_{0,\kappa})^2(q^2+\bar M^2_\kappa)^{D}}\right.\nonumber\\
& & \hspace{4.5cm}+\,T(\kappa\cdot\Delta r)\left(\frac{\bar \omega_q^\kappa}{q^2+\bar M^2_{0,\kappa}}-D\frac{\bar \omega_q^\kappa}{q^2+\bar M^2_\kappa}\right)\nonumber\\
& & \hspace{2.0cm}\left.-\,\frac{1}{2}T^2(\kappa\cdot\Delta r)^2\left(\frac{(\bar\omega_q^\kappa)^2}{(q^2+\bar M^2_{0,\kappa})^2}+\frac{(\bar \omega_q^\kappa)^2}{(q^2+\bar M^2_{0,\kappa})(q^2+\bar M^2_\kappa)}-2D\frac{(\bar\omega_q^\kappa)^2}{(q^2+\bar M^2_\kappa)^2}+\frac{D-1}{q^2+\bar M^2_{\kappa}}\right)\right].
\eeq
Then, using the basic $D$-dimensional integrals
\beq
\int \frac{d^Dq}{(2\pi)^D}\frac{1}{(q^2+M^2)^\alpha} & = & \frac{(M^2)^{D/2-\alpha}}{(4\pi)^{D/2}}\frac{\Gamma(\alpha-D/2)}{\Gamma(\alpha)}\,,\\
\int \frac{d^Dq}{(2\pi)^D}\ln(q^2+M^2) & = & -\frac{(M^2)^{D/2}}{(4\pi)^{D/2}}\Gamma(-D/2)\,,
\eeq
we find\footnote{The real part originates from the remark below Eq.~(\ref{eq:final}). The other terms do not require an explicit real part.}
\beq
& &  \delta V_{\bar r}(r)=\frac{T}{(4\pi)^{D/2}}\sum_\kappa\sum_{q\in\mathds{Z}}\Bigg[\frac{1}{2}\Big(2\,{\rm Re}\,(M^2_{0,\kappa})^{D/2}+D(\bar M^2_{\kappa})^{D/2}-(\bar M^2_{0,\kappa})^{D/2}-(D-1)(M^2_{\kappa})^{D/2}\nonumber\\
& & \hspace{5.0cm}-\,(M^2_{\kappa,+})^{D/2}-(M^2_{\kappa,-})^{D/2}\Big)\Gamma\left(-D/2\right)\nonumber\\
& & \hspace{4.5cm} +\,T(\kappa\cdot\Delta r)\,\bar\omega_q^\kappa\Big((\bar M^2_{0,\kappa})^{D/2-1}-D(\bar M^2_\kappa)^{D/2-1}\Big)\Gamma\left(1-D/2\right)\nonumber\\
& & \hspace{4.5cm} +\,\frac{1}{2}T^2(\kappa\cdot\Delta r)^2(\bar\omega_q^\kappa)^2\Big(2D(\bar M^2_\kappa)^{D/2-2}-(\bar M^2_{0,\kappa})^{D/2-2}\Big)\Gamma\left(2-D/2\right)\nonumber\\
& & \hspace{4.5cm} +\,\frac{1}{2}T^2(\kappa\cdot\Delta r)^2(\bar\omega_q^\kappa)^2\frac{(\bar M^2_{\kappa})^{D/2-1}-(\bar M^2_{0,\kappa})^{D/2-1}}{\bar M^2_\kappa-\bar M^2_{0,\kappa}}\Gamma\left(1-D/2\right)\nonumber\\
& & \hspace{4.5cm} -\,\frac{1}{2}T^2(\kappa\cdot\Delta r)^2(\bar M^2_\kappa)^{D/2-1}(D-1)\Gamma\left(1-D/2\right)\Bigg].
\label{eq:final_sum}
\eeq
\end{widetext}

\subsection{Taking the $\smash{\epsilon\to 0}$ limit}
We have argued above that $\delta V_{\bar r}(r)$ is UV-finite so we can, in principle, take the $\smash{\epsilon\to 0}$ limit without encoun-

\noindent{tering any divergence. The latter limit is tricky, however, because, in the present case, it does not commute with the Matsubara summation, see Ref.~\cite{vanEgmond:2022nuo} for a thorough discussion including additional examples.}

To see where the problem originates from, let us study the behavior of the dimensionally regularized summand in Eq.~(\ref{eq:final_sum}) for large values of $|\omega_q|$. One finds
\beq
& & -\frac{T(\kappa\cdot\Delta r)^3}{3(4\pi)^{D/2}}(D-1)^2\Gamma(2-D/2)|\omega_q|^D\nonumber\\
& & \hspace{0.5cm}\times\,\left[\frac{T^3}{\omega_q^3}+(D-3)\kappa\cdot\left(\bar r+\frac{\Delta r}{4}\right)\frac{T^4}{\omega_q^4}+\cdots\right].\label{eq:asym}
\eeq
From the point of view of power counting, both terms contribute a divergence to the corresponding Matsubara sum which, of course, seems contradictory with the fact that $\delta V_{\bar r}(r)$ is finite. However, the contribution from the first term of the expansion (\ref{eq:asym}) cancels when one considers both limits $\smash{\omega_q\to+\infty}$ and $\smash{\omega_q\to -\infty}$. Second, the next term in the expansion includes a factor $\smash{D-3=-2\epsilon}$ which turns the divergent sum $\sum_{q\in\mathds{Z^*}} 1/|\omega_q|^{1+2\epsilon}\sim 1/\epsilon\times 1/(2\pi T)^{1+2\epsilon}$ into a finite result. It follows that $\delta V_{\bar r}(r)$ is indeed finite but that one could miss a finite contribution if the $\smash{\epsilon\to 0}$ limit was taken too early, that is before the Matsubara summation. 

One way to proceed is to take the limit of the summand anyway, while introducing the finite contribution by hand. This method was applied for instance in Ref.~\cite{vanEgmond:2022nuo} for the evaluation of the gluon propagator in the center-symmetric Landau gauge. Here, we shall proceed in a slightly different (but equivalent) way that allows one to avoid some subtleties that were not discussed in Ref.~\cite{vanEgmond:2022nuo}.\footnote{In App.~\ref{app:eps}, for completeness, we shall apply the method of Ref.~\cite{vanEgmond:2022nuo} to the evaluation of the potential and discuss these subtleties in more detail.} 

First, noticing that the summand is actually a function of $\bar\omega^\kappa_q$ and $\Delta r$, we consider the first two terms of the asymptotic expansion as $|\bar\omega^\kappa_q|\to \infty$:
\beq
& & -\frac{T(\kappa\cdot\Delta r)^3}{3(4\pi)^{D/2}}(D-1)^2\Gamma(2-D/2)|\bar\omega^\kappa_q|^D\nonumber\\
& & \hspace{0.5cm}\times\,\left[\frac{T^3}{(\bar\omega^\kappa_q)^3}+(D-3)\frac{\kappa\cdot\Delta r}{4}\frac{T^4}{(\bar\omega_q^\kappa)^4}\right],
\eeq
and subtract them from the summand, which allows one to safely take the limit  $\smash{\epsilon\to 0}$ because the corresponding sum is now absolutely convergent. 

The subtracted contributions are finally added back in the form of additional series which can be expressed in terms of the Hurwitz zeta function
\beq
\zeta(s,z)\equiv\sum_{q=0}^\infty \frac{1}{(q+z)^s}\,.
\eeq
More specifically, the needed series are (the power of $2\pi T$ is introduced for convenience)
\beq
(2\pi T)^{n+2\epsilon}\sum_{q\in\mathds{Z}}\frac{|\bar\omega^\kappa_q|^{3-2\epsilon}}{(\bar\omega_q^\kappa)^{3+n}} & = & \sum_{q\in\mathds{Z}}\frac{\left|q+\frac{\kappa\cdot\bar r}{2\pi}\right|^{3-2\epsilon}}{\left(q+\frac{\kappa\cdot\bar r}{2\pi}\right)^{3+n}}\,,
\eeq

\vglue4mm

\noindent{with $\smash{n=0}$ or $\smash{n=1}$. Given $x\neq\mathds{Z}$, let us denote by $\{x\}$ the unique real number between $0$ and $1$ such that $x-\{x\}$ is an integer. We then have}
\beq
& & (2\pi T)^{n+2\epsilon}\sum_{q\in\mathds{Z}}\frac{|\bar\omega^\kappa_q|^{3-2\epsilon}}{(\bar\omega_q^\kappa)^{3+n}}\\\nonumber\\
& & \hspace{0.2cm}=\,\sum_{q=0}^\infty\frac{1}{\left(q+\left\{\frac{\kappa\cdot\bar r}{2\pi}\right\}\right)^{n+2\epsilon}}+\sum_{q=-\infty}^{-1}\frac{\left(-q-\left\{\frac{\kappa\cdot\bar r}{2\pi}\right\}\right)^{3-2\epsilon}}{\left(q+\left\{\frac{\kappa\cdot\bar r}{2\pi}\right\}\right)^{3+n}}\nonumber\\
& & \hspace{0.2cm}=\,\sum_{q=0}^\infty\frac{1}{\left(q+\left\{\frac{\kappa\cdot\bar r}{2\pi}\right\}\right)^{n+2\epsilon}}-(-1)^n\sum_{q=0}^\infty\frac{1}{\left(q+1-\left\{\frac{\kappa\cdot\bar r}{2\pi}\right\}\right)^{n+2\epsilon}}\nonumber\\
& & \hspace{0.2cm}=\,\zeta\left(n+2\epsilon,\left\{\frac{\kappa\cdot\bar r}{2\pi}\right\}\right)-(-1)^n\zeta\left(n+2\epsilon,1-\left\{\frac{\kappa\cdot\bar r}{2\pi}\right\}\right).\label{eq:thesum}\nonumber
\eeq
For $\smash{n=0}$, we use
 \beq
\zeta(2\epsilon,z)=-B_{1}(z)+{\cal O}(\epsilon)\,,
\eeq
where $\smash{B_1(z)=z-1/2}$ denotes the Bernoulli polynomial of order $1$. The sum (\ref{eq:thesum}) then gives
\beq
1-2\left\{\frac{\kappa\cdot\bar r}{2\pi}\right\}.
\eeq
For $\smash{n=1}$, we use
\beq
\zeta(1+2\epsilon,z)=\frac{1}{2\epsilon}-\psi(z)+{\cal O}(\epsilon)\,,
\eeq
where $\psi(z)$ is the digamma function. The sum (\ref{eq:thesum}) gives in this case
\beq
\frac{1}{\epsilon}-\psi\left(\left\{\frac{\kappa\cdot\bar r}{2\pi}\right\}\right)-\psi\left(1-\left\{\frac{\kappa\cdot\bar r}{2\pi}\right\}\right).
\eeq
Putting all the pieces together, we arrive at
\begin{widetext}
\beq
\delta V_{\bar r}(r) & = & \frac{T^4}{6\pi}\sum_\kappa(\kappa\cdot\Delta r)^3\left(2\left\{\frac{\kappa\cdot\bar r}{2\pi}\right\}-1+\frac{\kappa\cdot\Delta r}{4\pi}\right)\nonumber\\
& + & \frac{T}{\pi}\sum_\kappa\sum_{q\in\mathds{Z}}\Bigg[\frac{1}{12}\Big(2\,{\rm Re}\,(M^2_{0,\kappa})^{3/2}+3(\bar M^2_{\kappa})^{3/2}-(\bar M^2_{0,\kappa})^{3/2}-2(M^2_{\kappa})^{3/2}-(M^2_{\kappa,+})^{3/2}-(M^2_{\kappa,-})^{3/2}\Big)\nonumber\\
& & \hspace{2.0cm} -\,\frac{T}{4}(\kappa\cdot\Delta r)\,\bar\omega_q^\kappa\Big((\bar M^2_{0,\kappa})^{1/2}-3(\bar M^2_\kappa)^{1/2}\Big)+\frac{T^2}{4}(\kappa\cdot\Delta r)^2(\bar M^2_\kappa)^{1/2}+\frac{T^3}{6}(\kappa\cdot\Delta r)^3{\rm sgn}(\bar\omega^\kappa_q)\nonumber\\
& & \hspace{2.0cm} +\,\frac{T^2}{16}(\kappa\cdot\Delta r)^2(\bar\omega_q^\kappa)^2\left(\frac{6}{(\bar M^2_\kappa)^{1/2}}-\frac{1}{(\bar M^2_{0,\kappa})^{1/2}}-\frac{2}{(\bar M^2_{0,\kappa})^{1/2}+(\bar M^2_\kappa)^{1/2}}\right)\Bigg].\label{eq:V_final}
\eeq
\end{widetext}
We mention that both the sum and the correction terms in the first line obey the various symmetry identities discussed in Sec.~\ref{sec:sym}. This is one advantage with respect to the result that one obtains using the method of Ref.~\cite{vanEgmond:2022nuo} in which case, some of the symmetries reemerge only after one has added all contributions, see App.~\ref{app:eps}. We also mention that we could further improve the evaluation of the Matsubara sum by subtracting and adding back higher terms of the asymptotic expansion. The subtracted sum would converge faster and faster while the added terms can all be expressed in terms of the Hurwitz zeta function.\footnote{For the present application, such type of improvement is not really needed as we find that a few Matsubara sums ($|q|\lesssim 3$) give already a pretty accurate account of the potential in the relevant range of temperatures.}

 \subsection{Final expression}
 At the end of the day, the final result for the potential is
\begin{widetext}
\beq
V_{\bar r}(r) & = & V_{\bar r}(\bar r)+\frac{T}{2\pi^2}\sum_\kappa (\kappa\cdot\Delta r) \int_0^\infty \!\!\!dq\,q^2\,{\rm Im}\big[3n_{\bar\varepsilon^\kappa_q}-n_{\bar q^\kappa}\big]+\frac{T^2}{2g^2}M^2_{T,jk}\Delta r^j\Delta r^k\nonumber\\
& + & \frac{T^4}{6\pi}\sum_\kappa(\kappa\cdot\Delta r)^3\left(2\left\{\frac{\kappa\cdot\bar r}{2\pi}\right\}-1+\frac{\kappa\cdot\Delta r}{4\pi}\right)\nonumber\\
& + & \frac{T}{\pi}\sum_\kappa\sum_{q\in\mathds{Z}}\Bigg[\frac{1}{12}\Big(2\,{\rm Re}\,(M^2_{0,\kappa})^{3/2}+3(\bar M^2_{\kappa})^{3/2}-(\bar M^2_{0,\kappa})^{3/2}-2(M^2_{\kappa})^{3/2}-(M^2_{\kappa,+})^{3/2}-(M^2_{\kappa,-})^{3/2}\Big)\nonumber\\
& & \hspace{2.0cm} -\,\frac{T}{4}(\kappa\cdot\Delta r)\,\bar\omega_q^\kappa\Big((\bar M^2_{0,\kappa})^{1/2}-3(\bar M^2_\kappa)^{1/2}\Big)+\frac{T^2}{4}(\kappa\cdot\Delta r)^2(\bar M^2_\kappa)^{1/2}+\frac{T^3}{6}(\kappa\cdot\Delta r)^3{\rm sgn}(\bar\omega^\kappa_q)\nonumber\\
& & \hspace{2.0cm} +\,\frac{T^2}{16}(\kappa\cdot\Delta r)^2(\bar\omega_q^\kappa)^2\left(\frac{6}{(\bar M^2_\kappa)^{1/2}}-\frac{1}{(\bar M^2_{0,\kappa})^{1/2}}-\frac{2}{(\bar M^2_{0,\kappa})^{1/2}+(\bar M^2_\kappa)^{1/2}}\right)\Bigg].\label{eq:V_final}
\eeq
\end{widetext}
Let us mention that the non-commutation of the $\smash{\epsilon\to 0}$ limit and the Matsubara summation, which leads to the modification of the summand and the correction term in Eq.~(\ref{eq:V_final}), can be traced back to the second term in Eq.~(\ref{eq:final}). Since this term is very easily evaluated using the contour deformation technique, another possibility is to apply our strategy only to the other terms. In this case, there is no correction term to be added when permuting the $\smash{\epsilon\to 0}$ limit and the Matsubara sums. We collect the details in App.~\ref{app:other}.\footnote{This is actually the way the potential was originally computed in Ref.~\cite{vanEgmond:2021jyx}. We have checked numerically that the two formulas lead to the same results.}

For completeness, we also provide expressions for the first and second derivatives of the potential. Those are useful when studying the transition. Introducing a slightly more compact notation $\smash{X_s\equiv(M^2_{s,\kappa})^{1/2}}$ and using that
\beq
\frac{\partial X_0^2}{\partial r^j} & = & \kappa^j T\bar\omega^\kappa_q\,,\\
\frac{\partial X^2}{\partial r^j} & = & 2\kappa^j T\omega^\kappa_q\,,\\
\frac{\partial X_\pm^2}{\partial r^j} & = & \kappa^j T\bar\omega^\kappa_q\left(1\pm \frac{m^2}{X_+^2-X_-^2}\right),
\eeq
we find
\begin{widetext}
\beq
\frac{\partial V_{\bar r}}{\partial r^j} & = & \frac{T}{2\pi^2}\sum_\kappa \kappa^j \int_0^\infty \!\!\!dq\,q^2\,{\rm Im}\big[3n_{\bar\varepsilon^\kappa_q}-n_{\bar q^\kappa}\big]+\frac{T^2}{g^2}M^2_{T,jk}\Delta r^k+\frac{T^4}{6\pi}\sum_\kappa\kappa^j(\kappa\cdot\Delta r)^2\left(6\left\{\frac{\kappa\cdot\bar r}{2\pi}\right\}-3+\frac{\kappa\cdot\Delta r}{\pi}\right)\nonumber\\
& + & \frac{T^2}{8\pi}\sum_\kappa\kappa^j\sum_{q\in\mathds{Z}}\Bigg[\bar\omega^\kappa_q\Bigg(2\,{\rm Re}\,X_0-2\bar X_0-4X+6\bar X-X_+-X_--\frac{m^2}{X_++X_-}\Bigg)+4T^2(\kappa\cdot\Delta r)^2{\rm sgn}(\bar\omega^\kappa_q)\nonumber\\
& & \hspace{3.0cm}+\,4T(\kappa\cdot\Delta r)\big(\bar X-X\big)+T(\kappa\cdot\Delta r)(\bar\omega_q^\kappa)^2\!\left(\frac{6}{\bar X}-\frac{1}{\bar X_0}-\frac{2}{\bar X_0+\bar X}\right)\Bigg].\label{eq:first}
\eeq
\end{widetext}
We note that, for $\smash{r=\bar r}$, the RHS should be given only by the first term, see the discussion above. Using that $\smash{X_0=\bar X_0=X_-\in\mathds{R}}$ and $\smash{X=\bar X=X_+}$ in this limit, and $X^2=X_0^2+m^2$, we find that the remaining terms combine into
\beq
X-X_0-\frac{m^2}{X+X_0}=0\,.\nonumber
\eeq
Similarly, after some algebra, one finds
\begin{widetext}
\beq
\frac{\partial^2 V_{\bar r}}{\partial r^j\partial r^k} & = & \frac{T^2}{g^2}M^2_{T,jk}+\frac{T^4}{6\pi}\sum_\kappa\kappa^j\kappa^k(\kappa\cdot\Delta r)\left(12\left\{\frac{\kappa\cdot\bar r}{2\pi}\right\}-6+3\frac{\kappa\cdot\Delta r}{\pi}\right)\nonumber\\
& + & \frac{T^3}{8\pi}\sum_\kappa\kappa^j\kappa^k\sum_{q\in\mathds{Z}}\Bigg[(\bar\omega_q^\kappa)^2\Bigg({\rm Re}\,\frac{1}{X_0}+\frac{6}{\bar X}-\frac{1}{\bar X_0}-\frac{2}{\bar X_0+\bar X}-\frac{X_++X_-}{2X_+X_-}\left(1-\frac{m^2}{(X_++X_-)^2}\right)^2\Bigg)\nonumber\\
& & \hspace{4.0cm}+\,4\big(\bar X-X\big)-4\frac{(\omega^\kappa_q)^2}{X}+8T(\kappa\cdot\Delta r)\,{\rm sgn}(\bar\omega^\kappa_q)\Bigg].\label{eq:second}
\eeq
\end{widetext}
Once again, we can check that, when $r=\bar r$, all terms cancel except for the first one. This is because the remaining terms combine into
\beq
& & \frac{2}{X}-\frac{2}{X+X_0}-\frac{X+X_0}{2XX_0}\left(1-\frac{m^2}{(X+X_0)^2}\right)^2=0\,.\nonumber
\eeq
We mention that the integral in Eqs.~(\ref{eq:V_final}) and (\ref{eq:first}) vanishes for $\smash{\bar r=\bar r_c}$, see App.~\ref{app:sums}.

\section{Renormalization group}\label{sec:schemes}

To finalize the evaluation of the potential, we need to fix the finite parts of the renormalization factors. This is done by specifying a renormalization scheme which, in turn, provides a specification of the renormalized parameters $m$ and $g$ at a given renormalization scale $\mu$.

In what follows, we compare two popular schemes in the framework of the CF model, the vanishing momentum scheme (VM for short) that was used in Ref.~\cite{vanEgmond:2021jyx} as well as the infrared safe scheme (IR-safe for short) introduced in Ref.~\cite{Tissier:2011ey}.\\

\subsection{Renormalization schemes}
The CF model exhibits a non-renormalization theorem that derives from the anti-ghost shift symmetry $\smash{\bar c\to\bar c+\bar\lambda}$. As a consequence, the combination $Z_{g^2}Z_aZ_c^2$ of renormalization factors is UV finite, and one can take the renormalization of the coupling as\footnote{Recall that we have also chosen our scheme such that $\smash{Z_{\bar A}=Z_{g^2}^{-1}}$. Therefore, $\smash{Z_{\bar A}=Z_aZ_c^2}$.}
\beq
Z_{g^2}=Z_a^{-1}Z_c^{-2}\,.\label{eq:Taylor}
\eeq
This will be a common feature of the two schemes considered in this work. In both of them, we shall also fix the ghost renormalization $Z_c$ from the condition
\beq
D^{-1}_{T=0}(q=\mu;\mu)=\mu^2\,,
\eeq
\vglue1mm
\noindent{on the renormalized, vacuum ghost two-point function $D_{T=0}(q,\mu)$. At one-loop order, this leads to}
\begin{widetext}
\beq
Z_c=1+\frac{g^2N}{64\pi^2}\left[3\left(\frac{1}{\epsilon}+\ln\frac{\bar\Lambda^2}{m^2}\right)+5+\frac{m^2}{\mu^2}+\frac{\mu^2}{m^2}\ln\frac{\mu^2}{m^2}-\frac{(\mu^2+m^2)^3}{\mu^4m^2}\ln\left(1+\frac{\mu^2}{m^2}\right)\right].
\eeq
\end{widetext}

Next, we notice that the one-loop potential involves the renormalization factors $Z_a$ and $Z_{m^2}$ in the particular combination $Z_a Z_{m^2}$. In the IR-safe scheme, this product is determined by exploiting yet another non-renormalization theorem related to a modified BRST symmetry present within the CF model which implies that the combination $Z_{m^2}Z_aZ_c$ of renormalization factors is finite. One can then set
\beq
Z_aZ_{m^2}=Z_c^{-1}\,.\label{eq:nrm}
\eeq
In contrast, in the VM scheme, the same combination is determined, instead, from the condition
\beq
G^{-1}_{T=0}(q=0;\mu)=m^2(\mu)\,,\label{eq:q0}
\eeq
on the renormalized, vacuum gluon two-point function $G_{T=0}(q;\mu)$. Notice that $G^{-1}_{T=0}(q=0;\mu)$ is nothing but $M^2_{T=0}$. Therefore, in this scheme, the RHS of Eq.~(\ref{eq:Mvac}) is nothing but the squared renormalized mass $m^2(\mu)$, without having to evaluate $Z_a Z_{m^2}$. For later use, we note nonetheless that
\beq
Z_aZ_{m^2}=1-\frac{3g^2N}{64\pi^2}\left[\frac{1}{\epsilon}+\ln\frac{\bar\Lambda^2}{m^2}+\frac{5}{6}\right],\label{eq:zazm2}
\eeq
\vglue1mm
\noindent{where we have expanded in $g^2$ to one-loop accuracy.}

Finally, in order to fully fix the parameters and also to implement the renormalization group flow, see the next section, we need to determine $Z_a$ and $Z_{m^2}$ separately. In both schemes, one uses the condition
\beq
G^{-1}_{T=0}(q=\mu;\mu)=\mu^2+m^2(\mu)\,,
\eeq
which allows one to determine $Z_a$ knowing $Z_a Z_{m^2}$, and then, to also deduce $Z_{m^2}$. At one-loop order, one finds
\begin{widetext}
\beq
Z_a & = & 1-\frac{g^2N}{64\pi^2}\left[\left(\frac{1}{\epsilon}+\ln\frac{\bar\Lambda^2}{m^2}\right)\left(-\frac{26}{3}+3\frac{m^2}{\mu^2}\right)-\frac{121}{9}+21\frac{m^2}{\mu^2}\right.\nonumber\\
& & \hspace{1.5cm}\,+\frac{1}{4}\left(1+\frac{4m^2}{\mu^2}\right)^{3/2}\left(\left(\frac{\mu^2}{m^2}\right)^2-20\frac{\mu^2}{m^2}+12\right)\ln\frac{\sqrt{1+\frac{4m^2}{\mu^2}}-1}{\sqrt{1+\frac{4m^2}{\mu^2}}+1}\nonumber\\
& & \hspace{1.5cm}+\,\frac{1}{2}\left(1+\frac{m^2}{\mu^2}\right)^3\left(\left(\frac{\mu^2}{m^2}\right)^2-10\frac{\mu^2}{m^2}+1\right)\ln\left(1+\frac{\mu^2}{m^2}\right)\nonumber\\
& & \hspace{1.5cm}+\,\left.\frac{1}{4}\left(2-\left(\frac{\mu^2}{m^2}\right)^2\right)\ln\frac{\mu^2}{m^2}-2\left(\frac{m^2}{\mu^2}\right)^2\right]-(Z_aZ_{m^2}-1)\frac{m^2}{\mu^2}\,.\label{eq:za}
\eeq
\vglue20mm
\end{widetext}

\subsection{Renormalization Group flow}
The renormalization procedure introduces a scale $\mu$ at which the renormalization conditions are imposed and which eventually replaces the regulating scale $\Lambda$ introduced earlier. The dependence on $\mu$ is spurious of course because physical observables should not depend on it. In practice, however, observables are evaluated to a certain degree of approximation, which typically introduces a spurious dependence on $\mu$, thus potentially hindering predictability. 

Interestingly, this a priori annoying feature can be turned into a test of the quality of the approximation. Indeed, the better the approximation, the less scale dependence should be present in the results. In this work, we shall test our one-loop calculation of the potential by studying the scale dependence of various physical observables related to the confinement/deconfinement transition, such as the transition temperature, the spinodal temperatures (in the case of a first-order transition), or various order parameters for center-symmetry, see Sec.~\ref{sec:results} for more details.

For the test to make sense, as the scale $\mu$ is varied, the renormalized parameters $\smash{x=g^2}$ or $m^2$ should be varied along a renormalization group trajectory or ``line of constant physics''. This variation is encoded within the beta functions
\beq
\beta_x\equiv\mu\frac{dx}{d\mu}\,,
\eeq
where it is implicitly understood that the $\mu$-derivative needs to be taken at fixed bare parameters.

The beta functions can themselves be extracted from the knowledge of the corresponding renormalization factors $Z_x$. Indeed, since the bare parameters $Z_x x$ know nothing about the scale $\mu$, one has
\beq
\mu\frac{dZ_x}{d\mu}x+Z_x \mu\frac{dx}{d\mu}=0\,,
\eeq
and thus
\beq
\beta_x=-\mu\frac{d\ln Z_x}{d\mu}x\equiv-x\gamma_x\,,
\eeq
with $\gamma_x$ the anomalous dimension associated to the parameter $x$. 

Similarly, one associates anomalous dimensions $\gamma_y$ to the various field renormalization factors, with $\smash{y=a}$ or $c$. In both schemes considered in this work, the  $\gamma_x$'s and thus the $\beta_x$'s can be expressed solely in terms of the $\gamma_y$'s. First, owing to the condition (\ref{eq:Taylor}), one has $\gamma_{g^2}=-\gamma_a-2\gamma_c$ from which one deduces that
\beq
    \beta_{g^2}=g^2\left(\gamma_a+2\gamma_c\right).
\eeq
Moreover, in the IR-safe scheme, the condition (\ref{eq:nrm}) implies $\gamma_{m^2}=-\gamma_a-\gamma_c$, from which one deduces that
\beq
\beta_{m^2}=m^2\left(\gamma_a+\gamma_c\right).
\eeq
In contrast, in the VM scheme, from Eq.~(\ref{eq:zazm2}) and neglecting higher-order contributions in the flow, one obtains $\gamma_{m^2}=-\gamma_a$ and thus
\beq
\beta_{m^2}=m^2\gamma_a\,,
\eeq
which does not involve $\gamma_c$ in this case.

The anomalous dimensions $\gamma_c$ and $\gamma_a$ can be determined from the expressions for the renormalization constants $Z_c$ and $Z_a$ given above. The ghost renormalization is fixed equally in both schemes resulting in the same anomalous dimension
\begin{widetext}
\beq
\gamma_c & \! =& \! -\frac{g^2N}{32\pi^2t^2}\left[ 2t^2+2t-t^3\ln t+(t+1)^2(t-2)\ln(t+1) \right],
\eeq
where for brevity we have defined the dimensionless ratio $\smash{t\equiv\mu^2/m^2}$. As for the gluon anomalous dimension, its expression depends on the considered scheme, as seen from the differently defined product $Z_aZ_{m^2}$ which enters the expression (\ref{eq:za}) for $Z_a$. In the IR safe scheme, it reads \cite{Tissier:2011ey}
\beq
\gamma_a & \!=& \! -\frac{g^2N}{96\pi^2t^3}\bigg[ 17t^3-74t^2+12t-t^5\ln t+(t+1)^2(t-2)^2(2t-3)\ln(t+1)\nonumber \\
        &&\phantom{+++++}+t^{\frac{3}{2}}\sqrt{t+4}(t^3-9t^2+20t-36)\ln\left(\frac{\sqrt{t+4}-\sqrt{t}}{\sqrt{t+4}+\sqrt{t}}\right) \bigg],
\eeq
while in the VM scheme, one finds \cite{Tissier:2011ey}
\beq
\gamma_a & \!=& \! -\frac{g^2N}{96\pi^2t^3} \bigg[ 17t^3-\frac{175}{2}t^2+3t-t^5\ln t + (t+1)^2(2t^3-11t^2+20t-3)\ln(t+1) \nonumber \\
&&\phantom{++++++}+t^{\frac{3}{2}}\sqrt{t+4}(t^3-9t^2+20t-36)\ln\left(\frac{\sqrt{t+4}-\sqrt{t}}{\sqrt{t+4}+\sqrt{t}}\right) \bigg].
\eeq
\vglue8mm
\end{widetext}

\subsection{Initial conditions}
From the above expressions for the field anomalous dimensions, one can access the beta functions which, upon integration, give the runnings of the renormalized parameters in the considered scheme. 

Of course, the beta functions need to be integrated from a set of initial conditions specifying the values $m_0$ and $g_0$ of the renormalized parameters at a given scale $\mu_0$. Here, since the CF model is meant to be a phenomenological model for the Landau gauge-fixing in the infrared, we shall use the values obtained in Ref.~\cite{Tissier:2011ey,Barrios:2022zhu} by fitting the one-loop CF gluon and ghost vacuum two-point functions to the corresponding YM Landau gauge propagators as computed on the lattice.\footnote{Since the renormalization is done at $\smash{T=0}$, we can switch off the background for the present discussion.}

In the IR safe scheme, the fits are performed by setting the renormalization scale in the propagator expressions to $\smash{\mu=p}$ where $p$ is the momentum variable entering the propagators.\footnote{This is done to prevent the appearance of large logarithms in the UV tails.} This leads to the following values of the renormalized parameters at the scale $\smash{\mu_0=1}$ GeV:
\beq
& & \mbox{SU($2$):} \quad m_0=450\,{\rm MeV}, \quad g_0=5.2\,,\nonumber\\
& & \mbox{SU($3$):} \quad m_0=390\,{\rm MeV}, \quad g_0=3.7\,.\nonumber
\eeq
In the VM scheme, due to the presence of a Landau pole, the same fits were performed using $\smash{\mu=\sqrt{p^2+\alpha m^2_0}}$ with $\smash{\alpha=1}$ or $\smash{\alpha=2}$. One obtains two sets of initial conditions.\footnote{If no approximations were present, these two sets should correspond to the same renormalization group trajectory.} For $\smash{\alpha=1}$, one has
\beq
& & \mbox{SU($2$):} \quad m_0=600\,{\rm MeV}, \quad g_0=5.6\,,\nonumber\\
& & \mbox{SU($3$):} \quad m_0=500\,{\rm MeV}, \quad g_0=4.3\,,\nonumber
\eeq
whereas for $\smash{\alpha=2}$, one has\footnote{Some typos in Tables 5.5 and 5.6 of Ref.~\cite{Barrios:2022zhu} were pointed to us by the author.}
\beq
& & \mbox{SU($2$):} \quad m_0=600\,{\rm MeV}, \quad g_0=5.9\,,\nonumber\\
& & \mbox{SU($3$):} \quad m_0=500\,{\rm MeV}, \quad g_0=4.7\,.\nonumber
\eeq
Some other fits were performed in the VM scheme by fixing the renormalization scale in the propagator expressions to $\smash{\mu=\mu_0=1}$ GeV. The values of the parameters are in this case
\beq
& & \mbox{SU($2$):} \quad m_0=680\,{\rm MeV}, \quad g_0=7.5\,,\nonumber\\
& & \mbox{SU($3$):} \quad m_0=540\,{\rm MeV}, \quad g_0=4.9\,.\nonumber
\eeq
We have performed the now following analyses for this last choice as well, but noting that the results are essentially equal to those obtained using the other two VM scheme options and would only clutter up the figures we omit them from this work.

\begin{figure}[t]
    \centering
    \includegraphics[height=.23\textheight]{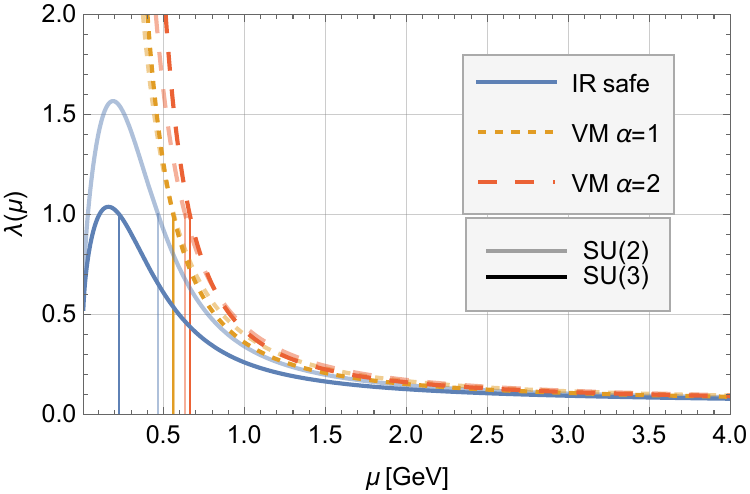}
    \caption{Running of the expansion parameter $\smash{\lambda={g^2N}/{16\pi^2}}$ in the considered renormalization schemes for SU(2) (transparent) and SU(3) (dark). The thin vertical lines mark the values of the renormalization scale at which this parameter goes above $1$.}
    \label{fig: coupling}
\end{figure}

Upon integrating the beta functions from the initial conditions given above, we obtain the running coupling and the running gluon mass. Figure \ref{fig: coupling} shows the running of the expansion parameter $\smash{\lambda\equiv g^2N/16\pi^2}$ which tells us about the range where our perturbative calculation is expected to be valid. In both schemes, $\lambda$ is perturbative ($\smash{\lambda<1}$) for intermediate to large values of the renormalization scale, as it should, and becomes larger than $1$ at the values marked with vertical lines in Fig.~\ref{fig: coupling}. As we will see in the next section, the region of interest always lies above this threshold, but we may also consider certain values around the region $\smash{\lambda\sim 1}$. For lower scales, the VM scheme exhibits an infrared Landau pole at which $\lambda$ diverges. In contrast, $\lambda$ remains finite in the IR safe scheme, vanishing for $\mu\to 0$.  Note that the drop-off from the Landau pole to perturbative values happens quickly, meaning we can have a valid perturbative expansion not too far from the Landau pole.

\section{Results}\label{sec:results}

In this section, we present our results in the center-symmetric Landau gauge $\smash{\bar r=\bar r_c}$.

\subsection{SU($2$) transition}

\begin{center}
\begin{figure}[t]
\includegraphics[height=0.23\textheight]{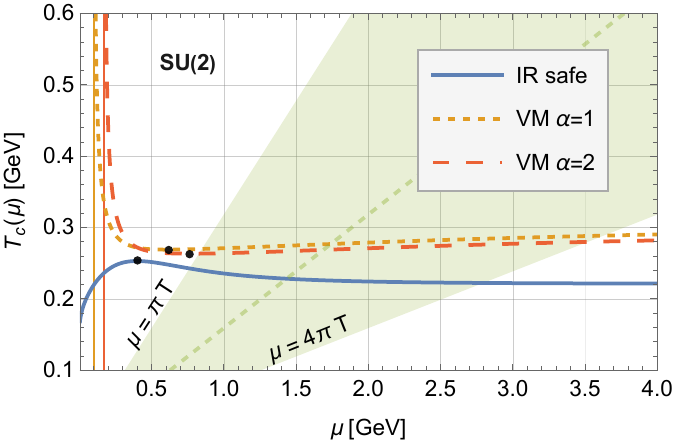}
\caption{SU($2$) transition temperature as a function of the renormalization scale $\mu$ in both the IR-safe and VM renormalization schemes. The conic band represents the region $\smash{\mu\in [\pi T,4\pi T]}$, with the dashed line representing the central value $\smash{\mu=2\pi T}$. The black dots correspond to the values obtained from a ``minimum sensitivity'' principle $\smash{dT_c/d\mu=0}$.}
\label{fig:Tc_SU2}
\end{figure}
\end{center}

In the SU($2$) case, the transition is continuous \cite{Gavai:1983av}. This means that the transition temperature can be extracted from the vanishing of the curvature of the potential at the center-symmetric point $\smash{r=\bar r_c=\pi}$. The latter is nothing but the squared mass $M^2_T$ introduced in Eq.~(\ref{eq:MT2}), which is just a number in the present case.  Since there are only two roots $\smash{\alpha=\pm 1}$ that contribute identically, and because $\smash{n_{\varepsilon-i\pi T}=-f_\varepsilon}$ with $\smash{f_\varepsilon\equiv 1/(e^{\beta\varepsilon}+1)}$ the Fermi-Dirac distribution, the condition determining the transition temperature reads
\beq
  0 & = & M^2_{T=0}+\frac{g^2T^2}{24}\left(1+\frac{7\pi^2}{5}\frac{T^2}{m^2}\right)\nonumber\\
    & - & \frac{g^2}{\pi^2}\int_0^\infty \!\!\!dq\,q^2\,\left(3\frac{m^2}{q^2}+6+\frac{q^2}{m^2}\right)\frac{f_{\varepsilon_q}}{\varepsilon_q}\,.
\eeq
Solving this equation for $T$ as a function of $\mu$, we obtain the curves $T_c(\mu)$ shown in Fig.~\ref{fig:Tc_SU2}.

Of course, the transition temperature is an observable and, as such, should not depend on the renormalization scale $\mu$.\footnote{This can be easily seen from the fact that a change of scheme leads to a multiplicative and temperature-independent, finite renormalization of the curvature mass, which thus does not affect the temperature at which the curvature mass vanishes. Of course, in practice, the multiplicative renormalization is not exactly satisfied, which introduces a spurious renormalization scale dependence to the transition temperature.} As already mentioned, however, approximations introduce a spurious scale dependence which can be used to test the quality of the approximation. For the test to make sense, it should be performed over a finite interval of values of $\mu$ to avoid a too-large separation between $\mu$ and the relevant scales of the problem that would invalidate the use of perturbation theory. 

Here, we use the conventional range $\smash{\mu\in [\pi T,4\pi T]}$, centered around $\smash{\mu=\omega_1=2\pi T}$, a value corresponding to the typical frequency scale associated with the temperature, the first Matsubara frequency. In Fig.~\ref{fig:Tc_SU2}, this interval corresponds to the highlighted, conic area. Similarly, the considered range should lie far away from possible Landau poles which can appear in certain renormalization schemes, of which the VM scheme is an example. The Landau pole is represented in Fig.~\ref{fig:Tc_SU2} by a vertical line.

We collect the corresponding temperature and renormalization scale ranges in physical units in Table \ref{tab: SU2 Tc table}. We observe that in both of the considered schemes, the scale dependence of $T_c$ is very mild,\footnote{This observation extends even further away from $\smash{\mu=4\pi T}$.} with a variation of $9\%$ and $6-7\%$ in the IR-safe and VM schemes respectively. These numbers are already quite good given the relatively simple one-loop approximation considered at this point. An even more stringent test would involve assessing whether and how much the renormalization scale dependence gets reduced as one includes the two-loop corrections. We leave this question for a future investigation. 

\begin{table}[h]
    \centering
    \begin{tabular}{r|cccl}
          &IR safe&  VM $\alpha=1$& VM $\alpha=2$ & \\ \hline
          $T_c\in$ & $[243,222]$ &  $[270,289]$ & $[264,280]$ & MeV \\
          $\mu\in$ & $[0.76,2.80]$ & $[0.85,3.63]$ & $[0.83,3.52]$ & GeV \\
    \end{tabular}
    \caption{The intervals for $T_c$ and $\mu$ with edge points satisfying $\mu=\pi T_c(\mu)$ and $\mu = 4\pi T_c(\mu)$ in the SU(2) case. Note that in the IR safe scheme $T_c(\mu)$ decreases over this interval.}
    \label{tab: SU2 Tc table}
\end{table}


Let us mention that, were we to take arbitrarily large values of $\mu$ we could obtain arbitrarily large values of $T_c$. To see this it is convenient to rescale all mass scales by $m$. The equation fixing $\smash{\tilde T_c\equiv T_c/m}$ becomes
\beq
  0 & = & \tilde M^2_{T=0}+\frac{g^2\tilde T^2}{24}\left(1+\frac{7\pi^2}{5}\tilde T^2\right)\nonumber\\
    & - & \frac{g^2}{\pi^2}\int_0^\infty \!\!\!dq\,q^2\,\left(\frac{3}{q^2}+6+q^2\right)\frac{\tilde f_{\tilde \varepsilon_q}}{\tilde\varepsilon_q}\,,\label{eq:two_res}
\eeq
with $\smash{\tilde f_{\varepsilon}\equiv 1/(e^{\varepsilon/\tilde T}+1)}$ and $\smash{\tilde\varepsilon_q\equiv\sqrt{q^2+1}}$, and where $\smash{\tilde M^2_{T=0}=1}$ in the VM scheme, and
\beq
\tilde M^2_{T=0} & = & 1-\frac{g^2}{32\pi^2}\left[\frac{5}{2}+\frac{m^2}{\mu^2}+\frac{\mu^2}{m^2}\ln\frac{\mu^2}{m^2}\right.\nonumber\\
& & \hspace{2.0cm}\left.-\,\frac{(\mu^2+m^2)^3}{\mu^4m^2}\ln\left(1+\frac{\mu^2}{m^2}\right)\right]\nonumber\\
\eeq
\begin{center}
\begin{figure}[t]
\includegraphics[height=0.23\textheight]{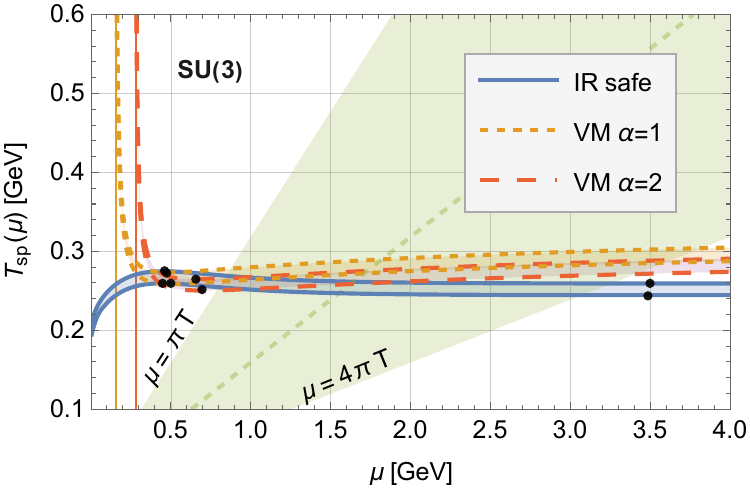}
\caption{SU($3$) higher spinodal temperatures as a function of the renormalization scale $\mu$ in both the IR-safe and VM renormalization schemes. For each, the colored band represents the temperature interval between the two spinodals. The conic band represents the region $\smash{\mu\in [\pi T,4\pi T]}$, with the dashed line representing the central value $\smash{\mu=2\pi T}$. The black dots correspond to the values obtained from a ``minimum sensitivity'' principle $\smash{dT_{\rm hsp}/d\mu=0}$.}
\label{fig:Tsp_SU3}
\end{figure}
\end{center}
in the IR-safe scheme. Since $g^2\sim (12/11)\pi^2/\ln (\mu/\mu_0)$ and $\tilde M^2_{T=0}$ goes to a constant in both schemes (either to $1$ or to $53/44$), we deduce that $\tilde T$ has to diverge. To see how it diverges, it is more convenient to rescale all masses by $T$ instead, in which case the equation reads
\beq
  0 & = & \tilde T^{-2}\tilde M^2_{T=0}+\frac{g^2}{24}\left(1+\frac{7\pi^2}{5}\tilde T^2\right)\nonumber\\
    & & -\,\frac{g^2}{\pi^2}\int_0^\infty \!\!\!dq\,q^2\,\left(\frac{3\tilde T^{-2}}{q^2}+6+q^2\tilde T^2\right)\frac{\tilde f_{\tilde \varepsilon_q}}{\tilde\varepsilon_q}\,,\nonumber\\
\eeq
with $\smash{\tilde f_{\varepsilon}\equiv 1/(e^{\varepsilon}+1)}$ and $\smash{\tilde\varepsilon_q\equiv\sqrt{q^2+\tilde T^{-2}}}$. At large $\tilde T$, it is found that the equation becomes $\smash{\tilde T^{-2}\tilde M^2_{T=0}\sim g^2/3}$ and thus $\smash{T^2\propto m^2/g^2}$. Since $m^2\propto (g^2)^{35/44}$ \cite{Tissier:2011ey,Reinosa:2017qtf}, we deduce that $T^2\propto (1/g^2)^{9/44}$ and thus diverges as $\smash{\mu\to\infty}$.

A similar behavior is observed when approaching the Landau pole (present in the VM scheme). Indeed, the coupling diverges in this limit which requires $\tilde T$ to approach one of the values that make the prefactor of $g^2$ in Eq.~(\ref{eq:two_res}) vanish. There are two such values $\tilde T=0$ and $\smash{\tilde T\equiv\tilde T_0\simeq 0.235}$. We have checked numerically that this second option is chosen, leading to $T\sim \tilde T_0 m$. Since $m$ diverges in the vicinity of the Landau pole,\footnote{We mention that, in the IR-safe, there also exist trajectories (not the ones considered here) that possess a Landau pole. But, for these trajectories, the mass vanishes at the Landau pole rather than diverging.} we deduce that the transition temperature diverges in this limit as well. 

Finally, in the IR-safe scheme, we can consider the limit $\mu\to 0$. Since the coupling goes to zero in this limit as well, and because $\tilde M^2_{T=0}$ approaches $1$, we deduce, once again that $\tilde T$ diverges like $1/g$, and thus $\smash{T\sim \sqrt{3}m/g}$. Now, in this scheme, $m/g$ goes to a constant \cite{Reinosa:2017qtf} and thus there is a limit to $T_c(\mu)$ as $\smash{\mu\to 0}$. In fact, since $\smash{m^2(\mu\to 0)/g^2(\mu\to 0)}$ can be directly related to the zero-momentum gluon propagator $G(0)$ in this scheme \cite{Tissier:2011ey}
\beq
\left.\frac{m^2(\mu)}{g^2(\mu)}\right|_{\mu\to 0}=G(p=0)\frac{m^4_0}{g_0^2}\,,
\eeq
one finds
\beq
T_c(\mu\to 0)=\sqrt{3G(0)}\frac{m^2_0}{g_0}\,.
\eeq
We stress once again that, even though we can analytically understand these limiting cases, the corresponding values of the renormalization scale should not be considered too seriously since they probably lie beyond the range of validity of a strict perturbative expansion.

Another popular choice of $\mu$ is the one that relies on the principle of minimal sensitivity. Since $T_c$ should be $\mu$-independent in the absence of approximations, one can define an optimal value $\mu^\star$ of the renormalization scale, and thus an optimal value $T_c^\star$ of the transition temperature, by enforcing the condition $\smash{dT_c/d\mu=0}$. We find $\smash{T_c^\star=253}$ MeV in the IR-safe scheme, corresponding to $\smash{\mu^\star/T_c^\star\simeq 0.50\pi}$, and $\smash{T_c^\star=269}$ or $263$ MeV (corresponding to $\smash{\alpha=1}$ or $2$) in the VM scheme, corresponding to $\smash{\mu^\star/T_c^\star\simeq 0.73\pi}$ or $0.92\pi$. In the case of the IR-safe scheme, we note that there is a second optimal value for $T_c$ but the renormalization scale ($\smash{\mu^\star/T_c^\star\simeq 7.3\pi}$) lies way above the interval $[\pi T,4\pi T]$.
Note that the lower, minimally sensitive $\mu$ for the IR-safe scheme is in the region where the coupling is not perturbative $\lambda(\mu^\star)\approx1.16$, even though close to $\lambda=1$.
Thus, this value should be taken with some care. Importantly, the range $\mu=\pi T_c\to4\pi T_c$ lies well within the perturbative region for all schemes.

So far our remarks concerned the internal consistency of the perturbative expansion within the CF model. One can also wonder to which extent this model allows one to capture non-trivial features of Yang-Mills theories. To this respect, let us note that the values obtained from the principle of minimal sensitivity in both schemes are quite close to each other, as they lie only $3-6\%$ apart. This is quite remarkable if one takes into consideration the fact that the initialization of the corresponding RG flows are external to the Curci-Ferrari model itself, since they are taken from fits to the Landau gauge correlators as computed on the lattice. Let us also stress that the predictions for $T_c$ within the CF are only $2-25\%$ below the value predicted by the simulations, $\smash{T_c^{\rm latt.}=295}$ MeV for the SU($2$) deconfinement transition. This observation will improve even further in the SU($3$) case.

\subsection{SU($3$) transition}
In the SU($3$) case, the curvature is a matrix with components corresponding to the color directions $3$ and $8$.  Within the center-symmetric gauge $\smash{r=\bar r_c=(4\pi/3,0)}$ and using charge conjugation invariance, one can argue that the transition occurs along the $\smash{r_8=0}$ direction. Consequently, the relevant quantity is the curvature along the color direction $3$, that is $M^2_{T,33}$, evaluated for $\smash{r=\bar r_c}$. 

It can again be argued that all roots contribute identically to $M^2_{T,33}$ and the condition for its vanishing eventually writes
\beq
    0 & = & M^2_{T=0}+\frac{g^2T^2}{24}\left(1+\frac{52\pi^2}{45}\frac{T^2}{m^2}\right)\nonumber\\
    & - & \frac{3g^2}{4\pi^2}\int_0^\infty \!\!\!dq\,\frac{q^2}{\varepsilon_q}\,\left(3\frac{m^2}{q^2}+6+\frac{q^2}{m^2}\right)\frac{e^{\beta\varepsilon_q}+2}{1+e^{\beta\varepsilon_q}+e^{2\beta\varepsilon_q}}\,.\nonumber\\
\eeq
Because the transition is first order in the SU($3$) case, this condition does not determine the transition temperature but, rather, the higher spinodal temperature $\smash{T_{\rm hsp}>T_c}$. However, this quantity should be renormalization scale independent as well in the absence of approximations, so studying its residual scale dependence is again a good way to test the quality of the one-loop approximation. 

\begin{center}
\begin{figure}[t]
\includegraphics[height=0.23\textheight]{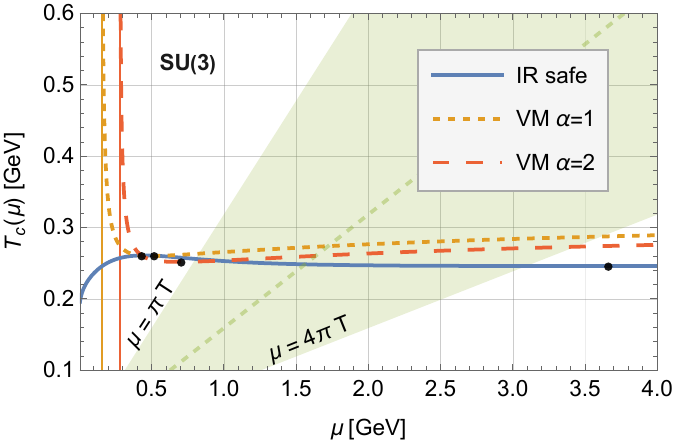}
\caption{SU($3$) transition temperature as a function of the renormalization scale $\mu$ in both the IR-safe and VM renormalization schemes. The conic band represents the region $\smash{\mu\in [\pi T,4\pi T]}$, with the dashed line representing the central value $\smash{\mu=2\pi T}$. The black dots correspond to the values obtained from a ``minimum sensitivity'' principle $\smash{dT_c/d\mu=0}$.}
\label{fig:Tc_SU3}
\end{figure}
\end{center}

Our results are collected in Fig.~\ref{fig:Tsp_SU3}. We observe similar features as for the transition temperature in the SU($2$) case, with the actual values falling between $\smash{259\ \rm{MeV}\lesssim T_{hsp}\lesssim304}$ MeV across all the schemes. The lower value corresponds to the minimum $T^\star_{\rm hsp}$ in the IR safe scheme and the higher value to the $\mu=4\pi T_{\rm hsp}(\mu)$ edge of the interval in the VM scheme with $\alpha=1$. A similar analysis has been
performed for the lower spinodal,\footnote{The lower spinodal does not occur at $\smash{r=\bar r_c}$, so the condition determining its location requires one to cancel both $\partial V_{\bar r_c}/\partial r^3$ and $\partial^2 V_{\bar r_c}/\partial (r^3)^2$, see Eqs.~(\ref{eq:first}) and (\ref{eq:second}). This leads to a system of equations determining both the value of the lower spinodal temperature and the value of $r$ at which it occurs.} with again very similar observations. Additionally, we observe that the spinodal temperatures lie rather close to the transition temperatures and could thus be used as a simpler alternative for it.

To  perform a faithful comparison with the SU($2$) case, we should access the actual transition temperature in the SU($3$) case. To do so, we need to locate the temperature (in-between the two spinodals) at which the two minima of the effective potential become degenerate. Once again, this quantity does not depend on the renormalization scheme since a change of scheme corresponds to a mere rescaling of the field variables that enter the effective potential. In practice, however, the transition temperature features a spurious scale dependence which we can use as a test of the quality of the one-loop approximations. Our results are presented in Fig.~\ref{fig:Tc_SU3}. The main features are similar to the ones in the previous plots, we highlight again the $[\pi T,4\pi T]$ interval as well as the extrema suggested by the principle of minimal sensitivity. We collect the temperature and renormalization scale ranges in Table \ref{tab: SU3 Tc table}.


As in the SU(2) case the scale dependence of $T_c$ over this interval is rather small, $4\%$ or $8-9\%$ in the IR safe or VM schemes respectively. One could again analyze whether these variations improve further when performing a two-loop calculation, which we leave for future investigation. Note that since the absolute values of $T_c$ went up/down and the length of the intervals down/up in the IR/VM scheme, the corresponding relative variation across the interval went down/up. This can partly be traced to the different initial conditions used for the renormalized parameters, e.g. the IR safe coupling $g_0$ decreased more from SU(2) to SU(3) compared to the VM schemes. We can again conclude that our perturbative expansion within the CF model seems reliable at one-loop order, since we obtained consistent results from different schemes. 

\begin{center}
\begin{figure}[t]
\includegraphics[height=0.23\textheight]{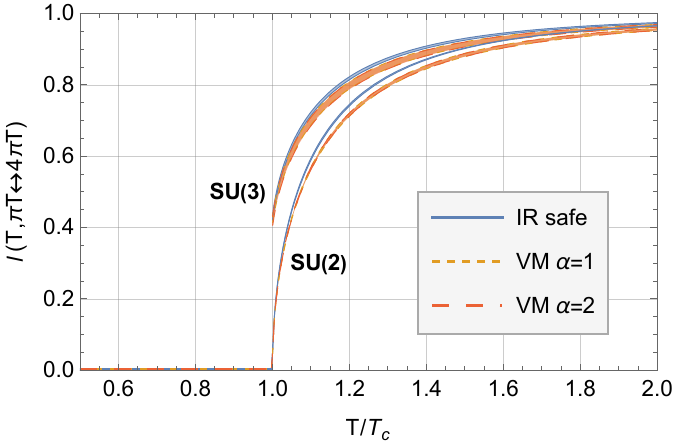}
\caption{SU(2) and SU(3) Polyakov loops as a function of the rescaled temperature $T/T_c$ and the renormalization scale $\smash{\mu\in[\pi T,4\pi T]}$. The regions spanned by taking $\mu$ between $\pi T$ and $4\pi T$ are shaded in.}
\label{fig: PolyakovLoop of T}
\end{figure}
\end{center}

Similar predictions arise using the principle of minimum sensitivity. In the IR safe scheme, both optimal values are now closer to the $[\pi T,4\pi T]$ range, giving $T_c^\star=261$ or $246$ MeV with $\mu^\star/T_c^\star\simeq 0.53\pi$ or $4.74\pi$. In the VM scheme the optimal temperatures are $T_c^\star=260$ or $252$ Mev with $\mu^\star/T_c^\star\simeq 0.63\pi$ or $0.89\pi$ (corresponding to $\smash{\alpha=1}$ or 2). Note that while all of these optimal points lie outside the respective $[\pi T,4\pi T]$ interval, due to the small dependence on the renormalization scale the ``optimal" temperatures do not differ greatly from the predictions given by the previous prescription. As for SU(2), there is one case where $T_c^\star$ is in the non-perturbative region, here for the VM $\alpha=1$ scheme, with $\lambda(\mu^\star)\approx1.18$. But again, the range $\mu=\pi T_c\to4\pi T_c$ lies well within the perturbative region for all schemes.


\begin{table}[h]
    \centering
    \begin{tabular}{r|cccl}
                   &IR safe.   &  VM $\alpha=1$ & VM $\alpha=2$ & \\ \hline
          $T_c\in$ & $[256,246]$ &  $[264,287]$ & $[252,273]$   & MeV \\
          $\mu\in$ & $[0.80,3.09]$ & $[0.83,3.61]$ & $[0.79,3.43]$ & GeV \\
    \end{tabular}
    \caption{The intervals for $T_c$ and $\mu$ with edge points satisfying $\mu=\pi T_c(\mu)$ and $\mu = 4\pi T_c(\mu)$ in the SU(3) case. Note that in the IR safe scheme $T_c(\mu)$ decreases over this interval.}
    \label{tab: SU3 Tc table}
\end{table}

We can also consider the variation with respect to the simulated value obtained from the lattice, $T_c^{\rm latt.}=270$ MeV. Our predictions for $T_c$ are now only $3-9\%$ below it in the IR scheme, or up to $6\%$ around it, including it in the interval, in both cases of the VM scheme. This is a significant improvement over the SU(2) case, where all schemes showed at least up to a $10\%$ deviation, see above. We presume that this improved behaviour can mostly be traced back to the fact that the fitted coupling constants are smaller in SU(3), rendering the loop expansion more accurate.

\vglue-10mm

\subsection{Order parameters}
The Polyakov loop \cite{Polyakov:1978vu,Svetitsky:1985ye,Pisarski:2002ji}
\beq
    \ell = \frac{1}{N}\left\langle\mathrm{tr}\,\mathcal{P}\exp\left(ig\int_0^\beta d\tau\,A_0^a(\tau,x)t^a\right)\right\rangle,
\eeq
is the most commonly used order parameter for the confinement/deconfinement transition and has been computed on the lattice, allowing for comparisons with continuum evaluations \cite{Herbst:2015ona,vanEgmond:2021jyx}. In this work, we have focused on an alternative quantity, the one-point function $\langle A\rangle_{\bar{A}_c}$ in the center-symmetric Landau gauge, as introduced in Section \ref{sec:framework}, or, equivalently, on the simpler quantity $r$, the argument of the effective potential. We can connect these quantities by noticing that $r$ appears implicitly in the definition of the Polyakov loop as it is contained in $A_0^a$. As discussed in Ref. \cite{vanEgmond:2021jyx}, at one-loop order, we can formally approximate the expression for the Polyakov loop by its tree level form, while plugging in the minimum of the potential $r_{\rm min}$ as found from the one-loop calculation.\footnote{Expanding the expression of the Polyakov loop to one-loop order and plugging in the minimum as found from the one-loop calculation would give a higher order correction.} We thus have
\beq
    \ell=\frac{1}{N}\mathrm{tr}\,e^{i r_{\rm min}^j t^j}+\mathcal{O}(g^4)\,,
\eeq
which plugging in the relevant generators and using only the $r^3$ component for SU(3) gives
\beq
    \mathrm{SU}(2):\quad\ell&=&\cos(r_{\rm min}/2)\,, \\
    \mathrm{SU}(3):\quad\ell&=&\frac{1}{3}\left(1+2\cos(r_{\rm min}/2)\right)\,.
\eeq
Note that plugging in the center-symmetric point $\pi$ or $4\pi/3$ respectively indeed gives zero.

The Polyakov loop usually gets renormalized which means that its renormalized version depends a priori on the renormalization scale. If the renormalization scheme is considered at zero temperature, however, the ratio of the Polyakov loop to the corresponding value at some reference temperature should be $\mu$-independent, again, up to truncation errors. In the present, one-loop, dimensionally regularized, continuum calculation, it turns out that the Polyakov loop does not require renormalization and thus, we expect the Polyakov loop itself to be quite independent of $\mu$, within the appropriate range.

One should bear in mind, however, that the transition temperature depends on $\mu$, as we saw in previous sections. Although small, this noticeable dependence affects the setting of the scale in physical units, and a faithful comparison of the Polyakov loop for various values of $\mu$ requires a rescaling of $T$ by $T_c$. In Fig. \ref{fig: PolyakovLoop of T} we show the SU(2) and SU(3) Polyakov loops as a function of this rescaled temperature and in the range between $\smash{\mu=\pi T}$ and $\smash{\mu=4\pi T}$ for each scheme. 

\begin{center}
\begin{figure}[b]
\includegraphics[height=0.23\textheight]{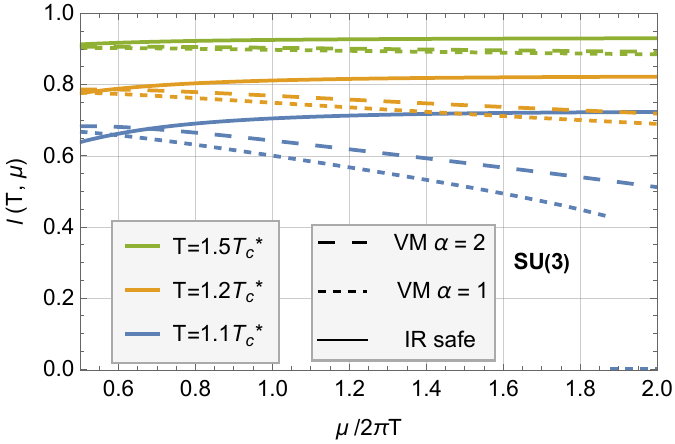}
\caption{SU(3) Polyakov loops as a function of renormalization scale $\mu$ in the region $\mu/(2\pi T)\in[0.5,2]$ for various temperatures, which are given in terms of the respective schemes minimally sensitive transition temperature $T_c^\star$.}
\label{fig: PolyakovLoop of mu}
\end{figure}
\end{center}

\vglue-8mm

Since the SU(2) transition is second-order, the Polyakov loop is continuous, while in the first-order SU(3) transition it has a discontinuity at the transition temperature. In both cases, after rescaling the temperature, only a very small dependence on the scheme and renormalization scale remains visible. The difference between the two versions of the VM scheme is negligible, while the IR safe scheme lies slightly above the others. Varying the renormalization scale in the range $\pi T\to4\pi T$ results in very small changes to the order parameter, which could only be visualized by using thinner lines for the edges of the interval and shading the inside, otherwise the variation is contained completely in the line thickness. We can conclude that most of the dependence on the renormalization scale is encoded in the transition temperatures, rather than in the order parameters.

Another way to verify the (small) effect of varying the renormalization scale $\mu$ is by instead evaluating the Polyakov loop at a fixed temperature and varying the renormalization scale. Looking at the SU(3) case in Fig.~\ref{fig: PolyakovLoop of T} we see that the variation due to the scale seems to be larger close to the transition temperature with the bands thinning as the temperature increases past $\sim1.5T_c$. In Fig.~\ref{fig: PolyakovLoop of mu} the results are shown for the fixed temperatures $T=aT_c^\star$, with $a\in\{1.1,1.2,1.5\}$ (bottom to top) and for renormalization scales in the range $[\pi T,4\pi T]$. In the IR safe scheme, the lower value of $T_c^\star$ was chosen to match the fact that it lies below the range, not above, as is the case for the VM schemes.

Using the optimal temperatures $T_c^\star$ found using the minimum sensitivity principle for each scheme allows in a sense to minimize the scheme dependence in the plot and focus on $\mu$-dependence, in contrast to taking the ratio with the $\mu$-dependent transition temperature as seen in Fig.~\ref{fig: PolyakovLoop of T}. This is confirmed by the fact that the values for the different schemes lie fairly close together for each temperature, even though the absolute temperatures $T$ used are different. While the lines are not completely flat, it is clear that across all schemes the dependence on the renormalization scale is small, and decreases further with increasing temperatures. We omit the analogous figure for SU(2), as the conclusions are the same, with the results showing a slightly stronger dependence on the scale $\mu$ especially for the IR safe scheme and low temperatures.

From the above analyses we conclude that the main source of the spurious dependencies on renormalization scheme and scale originate from the transition temperatures' dependencies, with the order parameters showing only very little additional variations after factoring this in. Thus to improve the results, an improvement of the calculation of the transition temperature is necessary, whether through higher loop orders, RG methods or others, and we expect the results for the Polyakov loop to follow suit.

\begin{center}
\begin{figure}[t]
\includegraphics[height=0.23\textheight]{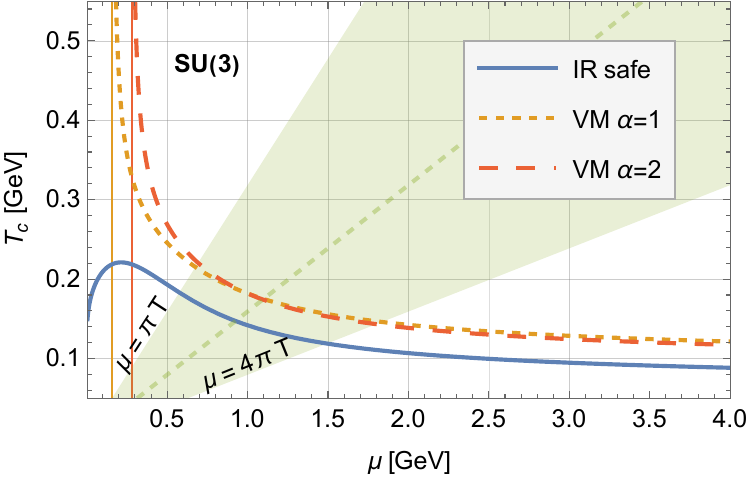}
\caption{SU($3$) transition temperature found using the background effective potential as a function of the renormalization scale $\mu$ in both the IR-safe and VM renormalization schemes. The conic band represents the region $\smash{\mu\in [\pi T,4\pi T]}$, with the dashed line representing the central value $\smash{\mu=2\pi T}$.}
\label{fig: background Tc SU(3)}
\end{figure}
\end{center}

\subsection{Background effective potential}\label{subsec: background pot}
We end this section by comparing our approach based on the center-symmetric Landau gauge to the one based on the minimization of the background effective potential $\smash{\tilde V({\bar r})\equiv V_{\bar r}(r=\bar r)}$. Although the latter is not a Legendre transform, it can be argued that, in the exact theory, minimizing $\tilde V(\bar r)$ is equivalent to minimizing $V_{\bar r_c}(r)$ \cite{vanEgmond:2023lnw}. However, this is not necessarily true anymore in the presence of approximations (like the one-loop approximation)  and/or modelling (such as the modelling of the gauge-fixing procedure considered in this paper), thus introducing a possible bias in the results. Our approach, being a true Legendre transform, should therefore be an improvement over the background effective potential.

At one-loop order, $\tilde V(\bar r)$ is given by \cite{Reinosa:2014ooa}
\begin{eqnarray}
 \tilde V({\bar r})=\frac{d-1}{2}\sum_\kappa\int_Q^T\ln\big[Q_\kappa^2+m^2\big]-\frac{1}{2}\sum_\kappa\int_Q^T\ln Q^2_\kappa\,.\label{eq:Vtilde}\nonumber\\
\end{eqnarray}
Except for a vacuum piece that we can disregard, there are no other divergences. The thermal piece can be evaluated using standard techniques to get
\begin{eqnarray}
 \tilde V({\bar r}) & \hat{=} & \frac{3}{4\pi^2}\sum_\kappa\int_0^\infty \!\!dq\,q^2\ln\Big(e^{2\beta\varepsilon_q}-2e^{\beta\varepsilon_q}\cos(\bar r\cdot\kappa)+1\Big)\nonumber\\
 & - & \frac{1}{4\pi^2}\sum_\kappa\int_0^\infty \!\! dq\,q^2\ln\Big(e^{2\beta q}-2e^{\beta q}\cos(\bar r\cdot\kappa)+1\Big)\,.\nonumber\\
\end{eqnarray}
One can also evaluate the first derivatives
\beq
\frac{\partial\tilde V}{\partial \bar r^j} & = & T\sum_\kappa\kappa^j\left[(d-1)\int_Q^T\frac{\omega^\kappa_q}{Q_\kappa^2+m^2}-\int_Q^T\frac{\omega^\kappa_q}{Q^2_\kappa}\right]\,,\nonumber\\
\eeq
and
\beq
\frac{\partial^2\tilde V}{\partial \bar r^j\partial\bar r^k} & = & T^2\sum_\kappa\kappa^j\kappa^k\left[-\int_Q^T\left(\frac{1}{Q_\kappa^2}-\frac{2(\omega^\kappa_q)^2}{(Q_\kappa^2)^2}\right)\right.\nonumber\\
& & \left.+\,(d-1)\int_Q^T\left(\frac{1}{Q_\kappa^2+m^2}-\frac{2(\omega^\kappa_q)^2}{(Q_\kappa^2+m^2)^2}\right)\right]\,.\nonumber\\
\eeq

Notice that the first derivative is the same as in Eq.~(\ref{eq:V1}) or in Eq.~(\ref{eq:V11}) but with $\bar r$ replaced by $r$. Moreover, It is easily checked using Eqs.~(\ref{eq:44})-(\ref{eq:45}) that the vacuum piece of the second derivative vanishes identically, in line with the fact that the vacuum piece of the potential (\ref{eq:Vtilde}) does not depend on $\bar r$. The second derivative is then just given by its thermal piece which can be computed using Eqs.~(\ref{eq:52})-(\ref{eq:53}). We find
\beq
\frac{\partial^2\tilde V}{\partial \bar r^j\partial\bar r^k} & \,\hat{=}\, & T^2\sum_\kappa\kappa^j\kappa^k\!\!\int_Q^T\!\!\left[3\left(2+\frac{m^2}{q^2}\right)\!\frac{1}{Q_\kappa^2+m^2}-\frac{2}{Q_\kappa^2}\right]\nonumber\\
& = & T^2\sum_\kappa\kappa^j\kappa^k\!\!\int_Q^T\!\!\left[3\left(2+\frac{m^2}{q^2}\right)\!\frac{n_{\varepsilon^\kappa_q}}{\varepsilon_q}-2\frac{n_{q^\kappa}}{q}\right].\nonumber\\
\eeq
Solving for the transition temperature can be done in analogy to the center-symmetric potential. In the SU(2) case sit suffices to find the temperature at which the curvature at the center-symmetric point vanishes, whereas this will again only give the upper spinodal temperature in the SU(3) case. There is an advantage over the center-symmetric potential though. Since there is no explicit dependence on the coupling in the potential and the only scale is given by the mass, everything can be solved in terms of $\smash{\Tilde{T}\equiv T/m}$, independent of the scheme. The running is implicit in $m(\mu)$ and thus the variations of $T_c$ with scale and scheme will come directly from the variations of the mass. We find that for SU(2) $\Tilde{T_c}\simeq0.336$ and for SU(3) $\Tilde{T_c}\simeq0.363$ with $r_{\rm min}(T_c+\epsilon)\simeq2.41$. For completeness, the higher and lower spinodal temperatures are $\Tilde{T}_{\rm hsp}\simeq0.38$ and $\Tilde{T}_{\rm lsp}\simeq0.361$ at $r_{\rm lsp}\simeq2.85$. At one-loop order, there appears also an artifact where the $r_{\rm min}$'s vanish exactly from a certain temperature on, corresponding to a maximal Polyakov loop or reaching a zero free energy requirement. This happens at $\Tilde{T}\simeq0.50$ in both cases, the defining equations for this temperature are equivalent in SU(2) and SU(3).

In Fig.~\ref{fig: background Tc SU(3)} we illustrate the type of scheme and scale dependence that we find using the background field effective potential. The transition temperatures vary much stronger with the scale compared to the ones computed using the center-symmetric potential, and are also generally further away from the lattice values. The variations across the $[\pi T,4\pi T]$ interval are of up to $53\%$ and the distance from the lattice value of $19-44\%$. The absolute values range between 119 MeV and 220 MeV. Additionally, the behaviour for large $\mu$ is also different, with the transition temperatures vanishing logarithmically, like the mass.

\section{Conclusions}
We have studied the confinement/deconfinement transition of pure Yang-Mills theories using perturbative methods. For this, we worked in the recently introduced center-symmetric Landau gauge with a Curci-Ferrari mass term modelling the effect of Gribov copies in the infrared. Extending previous work in this setup \cite{vanEgmond:2021jyx}, we present a comprehensive overview of the methods used to evaluate the one-loop potential for the one-point function, one of the possible order parameters for the confinement/deconfinement transition in this gauge. In this work, particular focus was placed on studying the renormalization scale and scheme dependence of various observables, this in view of both further testing the inner consistency of perturbative calculations within the CF model and checking the adequacy of the model as an effective and efficient description of background Landau-gauge YM theories in the infrared. 

We have shown that already the present one-loop order calculation is precise, with small dependencies on the renormalization scale and scheme in both the SU(2) and SU(3) cases. The transition temperature shows variations with the scale of under 9\%. Additionally, the results are accurate to the temperature found in lattice computation within 25\% for SU(2) and 9\% for SU(3), where the loop expansion is better behaved. In the SU(3) case we have also computed the lower and higher spinodal temperatures around the transition temperature and checked that they follow the same behavior considering the residual scale dependence. These results hold across the diverse renormalization schemes, further suggesting that the background gauge Curci-Ferrari model is a valid way of describing finite temperature Yang-Mills theory. While there is some visible variation with scale and scheme in the transition temperature, the order parameter shows barely any additional variations that cannot be traced back to the transition temperature. We have compared the current approach to previous methods involving a background effective potential and see a clear improvement in both the accuracy and precision of the transition temperatures. 

A natural continuation of this analysis is to compute the two-loop potential with the corresponding renormalization and check whether the results improve. Another extension is the application to QCD. In this case, it is well known that perturbative methods are inefficient, even within the Curci-Ferrari model. However, in this latter context, the good control of the pure gauge sector has allowed for the set-up of a controlled non-perturbative expansion scheme which we plan to investigate in the future. In this case as well, renormalization group effects need to be taken into account.

\acknowledgements{We thank Duifje van Egmond for helpful discussions and collaboration on related aspects. We also thank Nahuel Barrios for private communications regarding the setting of the parameters and the initialization of the RG flows.}

\appendix

\section{Determinant}\label{app:det}

As already explained, the one-loop effective potential is obtained from the determinant of the quadratic part of the action (\ref{eq:action}) expanded around a classical configuration of the fields. Recall that the classical ghost, anti-ghost, and Nakanishi-Lautrup fields can be taken equal to zero. For simplicity, the classical gauge field configuration is denoted $A_\mu$. This means that we have to expand the action $S_{\bar A}[A+a]$ in powers of the fluctuation $a$. 

From now on, we take both $\bar A$ and $A$ constant, temporal and abelian, see Eqs.~(\ref{eq:bg}) and (\ref{eq:res}).

\subsection{Quadratic Part}
In the ghost sector, the quadratic part is simply
\beq
-\int_x \bar c^a(x) \bar D_\mu^{ab} D_\mu^{bc}c^c(x)\,,
\eeq
where we have integrated by parts. In order to identify the quadratic part in the gluon sector, we note that
\beq
F^a_{\mu\nu}[A+a]=F^a_{\mu\nu}[A]+D^{ab}_\mu a^b_\nu-D^{ab}_\nu a^b_\mu+gf^{abc}a^b_\mu a^c_\nu\,,\nonumber\\
\eeq
where the first term in RHS vanishes in the present case. The quadratic part in the gluon sector then reads
\beq
& &\int_x\frac{1}{2}\Big(D^{ab}_\mu a^b_\nu(x)  D^{ac}_\mu a^c_\nu(x)-D^{ab}_\mu a^b_\nu(x) D^{ac}_\nu a^c_\mu(x)\nonumber\\
& & \hspace{1.0cm}+\,m^2(a_\mu^a(x))^2\Big)+\int_x ih^a(x)\bar D_\mu^{ab} a_\mu^b(x)\,.
\eeq
One should not forget the last term coupling $a_\mu$ to $h$ since it is also quadratic. After an integration by parts, we arrive at
\beq
& & \int_x\frac{1}{2}a^a_\mu(x)\Big(D^{ac}_\nu D^{cb}_\mu- \delta_{\mu\nu}(D^{ac}_\rho D^{cb}_\rho -m^2\delta^{ab})\Big)a^b_\nu(x)\nonumber\\
& & +\frac{1}{2}\int_x ih^a(x)\bar D_\nu^{ab} a_\nu^b(x)-\frac{1}{2}\int_x a_\mu^a(x) \bar D_\mu^{ab}(ih^b(x))\,.\label{eq:quad}\nonumber\\
\eeq

\subsection{Color Structure}
The index structure in Eq.~(\ref{eq:quad}) is intricate since it is neither diagonal in the Lorentz indices nor in the color indices. Let us first deal with the color structure.

Because both $A$ and $\bar A$ are constant, temporal and Abelian, it is convenient to switch to a Cartan-Weyl color basis. Indeed, in such a basis, the action of the covariant derivative writes
\beq
D_\mu(X^\kappa t^\kappa) & = & \partial_\mu(X^\kappa t^\kappa)-ig[A_\mu,X^\kappa t^\kappa]\nonumber\\
& = & (\partial_\mu X^\kappa) t^\kappa-iT\delta_{\mu0}\,rX^\kappa \left[\frac{\sigma_3}{2}, t^\kappa\right]\nonumber\\
& = & (\partial_\mu X^\kappa) t^\kappa-iT\delta_{\mu0}\,r \kappa X^\kappa  t^\kappa\nonumber\\
& = & (\partial_\mu-iT\delta_{\mu0}\,r \kappa) X^\kappa  t^\kappa\equiv (D_\mu^\kappa X^\kappa) t^\kappa\,,\nonumber\\
\eeq
and similarly of course for $\bar D_\mu(X^\kappa t^\kappa)$ with $r$ replaced by $\bar r$. Altogether, the quadratic part within a Cartan-Weyl color basis reads then
\beq
& & \int_x\frac{1}{2}a^{-\kappa}_\mu(x)\Big(D^\kappa_\nu D^\kappa_\mu- \delta_{\mu\nu}(D^\kappa_\rho D^\kappa_\rho -m^2)\Big)a^\kappa_\nu(x)\nonumber\\
& & +\frac{1}{2}\int_x ih^{-\kappa}(x)\bar D_\nu^\kappa a_\nu^\kappa(x)-\frac{1}{2}\int_x \bar  a_\mu^{-\kappa}(x) \bar D_\mu^\kappa (ih^\kappa(x))\,.\nonumber\\
\eeq
The presence of the labels $-\kappa$ relates to the fact that ${\rm tr}\,t^\kappa t^\lambda\propto\delta^{\kappa(-\lambda)}$ rather than ${\rm tr}\,t^a t^b\propto\delta^{ab}$ for the usual cartesian bases. This will make even more sense in the next section.

\subsection{Lorentz Structure}
It is more convenient to evaluate the determinant of the quadratic form in Fourier space. With the convention $\partial_\mu\to -iQ_\mu$, we find
\beq
& & \int_Q\frac{1}{2}a^{-\kappa}_\mu(-Q)\Big(\delta_{\mu\nu}(Q^\kappa_\rho Q^\kappa_\rho +m^2)-Q^\kappa_\nu Q^\kappa_\mu\Big)a^\kappa_\nu(Q)\nonumber\\
& & +\int_Q h^{-\kappa}(-Q)\bar Q_\nu^\kappa a_\nu^\kappa(Q)-\int_Q \bar  a_\mu^{-\kappa}(-Q) \bar Q_\mu^\kappa h^\kappa(Q)\,,\nonumber\\
\eeq
where $Q^\kappa_\mu$ and $\bar Q^\kappa_\mu$ are the generalized momenta introduced in the main text and which combine additively the momentum and the charge, which are both conserved. From this perspective, it seems natural that  $-\kappa$ appears in those Fourier components of momentum $-Q$.

In matrix form, the quadratic form in the gluon sector writes
\beq
\left(\begin{array}{cc}
\delta_{\mu\nu}(Q^\kappa_\rho Q^\kappa_\rho +m^2)-Q^\kappa_\mu Q^\kappa_\nu & -\bar Q^\kappa_\mu\\
\bar Q^\kappa_\nu & 0
\end{array}\right).\label{eq:quadro}
\eeq
To evaluate its determinant, we use the Schur decomposition of a square matrix made of four square blocks $A$, $B$, $C$ and $D$, with $A$ invertible:
\beq
\left(\begin{array}{cc}
A & B\\
C & D
\end{array}\right)=\left(\begin{array}{cc}
A & 0\\
C & 1
\end{array}\right)\left(\begin{array}{cc}
1 & A^{-1}B\\
0 & D-CA^{-1}B
\end{array}\right).
\eeq
From this decomposition, it follows that
\beq
{\rm det}\,\left(\begin{array}{cc}
A & B\\
C & D
\end{array}\right)={\rm det}\,A\times {\rm det}(D-CA^{-1}B)\,,
\eeq
where $D-CA^{-1}B$ is known as the Schur's complement.

In the present case, the block $A$ writes
\beq
A & = & \delta_{\mu\nu}(Q_\kappa^2 +m^2)-Q^\kappa_\mu Q^\kappa_\nu\nonumber\\
& = & (Q_\kappa^2 +m^2)P^\perp_{\mu\nu}(Q_\kappa)+m^2P^\parallel_{\mu\nu}(Q_\kappa)\,,
\eeq
where we have introduced the usual orthogonal projectors
\beq
P^\parallel_{\mu\nu}(Q_\kappa)\equiv\frac{Q^\kappa_\mu Q^\kappa_\nu}{Q^2_\kappa} \quad {\rm and} \quad P^\perp_{\mu\nu}(Q_\kappa)\equiv\delta_{\mu\nu}-\frac{Q^\kappa_\mu Q^\kappa_\nu}{Q^2_\kappa}\,,\nonumber\\
\eeq
but with respect to the generalized momentum $Q^\kappa_\mu$. The block $A$ is clearly invertible, with inverse $A^{-1}$ given by
\beq
A^{-1}=\frac{P^\perp_{\mu\nu}(Q_\kappa)}{Q_\kappa^2 +m^2}+\frac{P^\parallel_{\mu\nu}(Q_\kappa)}{m^2}
\eeq
and determinant
\beq
{\rm det}\,A=m^2(Q_\kappa^2 +m^2)^{d-1}\,,
\eeq
the power of $d-1$ coming from the fact that $P^\perp(Q_\kappa)$ corresponds to an eigenspace of dimension $d-1$. Finally, the Schur complement is just a number in this case and is computed to be
\beq
D-CA^{-1}B & = & \frac{\bar Q^\kappa_\mu P^\perp_{\mu\nu}(Q_\kappa)\bar Q^\kappa_\nu}{Q_\kappa^2 +m^2}+\frac{\bar Q^\kappa_\mu P^\parallel_{\mu\nu}(Q_\kappa)\bar Q^\kappa_\nu}{m^2}\nonumber\\
& = & \frac{\bar Q_\kappa^2}{Q_\kappa^2 +m^2}+\frac{(\bar Q_\kappa\cdot Q_\kappa)^2}{Q^2_\kappa}\left[\frac{1}{m^2}-\frac{1}{Q_\kappa^2+m^2}\right]\nonumber\\
& = & \frac{1}{Q_\kappa^2 +m^2}\left[\bar Q_\kappa^2+\frac{(\bar Q_\kappa\cdot Q_\kappa)^2}{m^2}\right].
\eeq
Here, we have introduced the notation $X^\kappa\cdot Y^\kappa\equiv X^\kappa_\mu Y^\kappa_\mu$ where a summation over $\mu$ is implied. Putting all the pieces together, the determinant of the quadratic form (\ref{eq:quadro}) writes
\beq
(Q_\kappa^2 +m^2)^{d-2}\left[m^2\bar Q_\kappa^2+(\bar Q_\kappa\cdot Q_\kappa)^2\right].\label{eq:det_g}
\eeq

\subsection{Final Result}
The logarithm of the determinant (\ref{eq:det_g}) contributes to the effective potential with a factor 
$1/2$. Similarly, the quadratic part in the ghost sector gives the determinant $\prod_Q \prod_\kappa Q_\kappa\cdot\bar Q_\kappa$ whose logarithm contributes to the effective potential with a factor $-1$. We mention that, if $\smash{Q=0}$ or $\smash{\kappa=0}$, then $\smash{Q_\kappa\cdot\bar Q_\kappa}$ is positive. Moreover, if $\smash{Q\neq 0}$ and $\smash{\kappa\neq0}$, to each pair $(Q,\kappa)$, we can associate a pair $(-Q,-\kappa)$ such that the combined contribution of these two modes to the determinant is positive
\beq
(Q_\kappa\cdot\bar Q_\kappa)((-Q)_{-\kappa}\cdot\overline{(-Q)}_{-\kappa})=(Q_\kappa\cdot\bar Q_\kappa)^2\,.
\eeq
All in all, this means that the ghost contribution writes$-(1/2)\ln (Q_\kappa\cdot\bar Q_\kappa)^2$. Combining this result with (\ref{eq:det_g}), we arrive at the formula given in the main text.

\section{Alternative approach}\label{app:other}
As suggested in the main text, an alternative approach to evaluate the potential is to rewrite it as
\beq
 V_{\bar r}(r)=\hat V_{\bar r}(r)+\frac{d-2}{2}\sum_\kappa\int_Q^T\ln\big[Q_\kappa^2+m^2\big],\label{eq:alternative}
\eeq
with
\begin{eqnarray}
\hat V_{\bar r}(r) & \equiv & \frac{m^2T^2}{2g^2}(r-\bar r)^2\nonumber\\
& + & \frac{1}{2}\sum_\kappa\int_Q^T\ln\left[1+\frac{m^2\bar Q_\kappa^2}{(Q_\kappa\cdot\bar Q_\kappa)^2}\right],
\end{eqnarray}
and to apply the strategy presented in the main text to $\hat V_{\bar r}(r)$ only, the explicit sum-integral in Eq.~(\ref{eq:alternative}) being computed as usual, via the analytic evaluation of the sum and the numerical evaluation of the resulting momentum integral. Then, we have to compute
\beq
\hat V_{\bar r}(r)=[\hat V_{\bar r}(r)]_2+\delta\hat V_{\bar r}(r)\,,
\eeq
with
\beq
\delta\hat V_{\bar r}(r)\equiv \hat V_{\bar r}(r)-[\hat V_{\bar r}(r)]_2\,.
\eeq
We find 
\beq
[\hat V_{\bar r}(r)]_2 & \equiv & \hat V_{\bar r}(\bar r)+\sum_j \left.\frac{\partial \hat V_{\bar r}(r)}{\partial r^j}\right|_{r=\bar r}\Delta r^j\nonumber\\
& + & \frac{1}{2} \sum_{j,k} \left.\frac{\partial^2 \hat V_{\bar r}(r)}{\partial r^j\partial r^k}\right|_{r=\bar r}\Delta r^j \Delta r^k\,,\label{eq:taylorr}
\eeq
with
\beq
\hat V_{\bar r}(\bar r)=\sum_\kappa\left[\frac{1}{2}\int_Q^T\ln\big(\bar Q_\kappa^2+m^2\big)-\frac{1}{2}\int_Q^T\ln \bar Q_\kappa^2\right],\label{eq:VV0}\nonumber\\
\eeq
\beq
\left.\frac{\partial \hat V_{\bar r}(r)}{\partial r^j}\right|_{r=\bar r}\!\!\!\!=T\sum_\kappa \kappa^j \left[\int_Q^T\!\frac{\bar\omega_q^\kappa}{\bar Q_\kappa^2+m^2}-\!\int_Q^T\!\frac{\bar\omega_q^\kappa}{\bar Q_\kappa^2}\right],\label{eq:VV1}\nonumber\\
\eeq
and
\beq
& & \left.\frac{\partial^2 \hat V_{\bar r}}{\partial r^j\partial r^k}\right|_{r=\bar r}\nonumber\\
& & \hspace{0.5cm}=\,T^2\Bigg[Z_a Z_{m^2}\frac{m^2}{g^2}\delta^{jk}+\sum_\kappa \kappa^j\kappa^k\nonumber\\
& & \hspace{0.7cm}\times\,\left( \int^T_Q \frac{(\bar\omega_q^\kappa)^2}{\bar Q_\kappa^4}-2\int^T_Q\frac{(\bar\omega_q^\kappa)^2}{(\bar Q_\kappa^2+m^2)^2}\right.\nonumber\\
& & \hspace{3.0cm}+\,\left.\int^T_Q\frac{(\bar\omega_q^\kappa)^2}{\bar Q^2_\kappa(\bar Q^2_\kappa+m^2)}\right)\Bigg].\label{eq:VV2}
\eeq
As before, we can ignore $\hat V_{\bar r}(\bar r)$. The first derivative is purely thermal and given by
\beq
\left.\frac{\partial \hat V_{\bar r}(r)}{\partial r^j}\right|_{r=\bar r}=\frac{T}{2\pi^2}\sum_\kappa \kappa^j \int_0^\infty \!\!\!dq\,q^2\,{\rm Im}\,\big[n_{\varepsilon^\kappa_q}-n_{q^\kappa}\big]\,.\nonumber\\
\eeq
As for the second derivative, it contains both a vacuum and a thermal part. Interestingly enough, the vacuum contribution is the same as before,\footnote{This is because, the vacuum contribution to the explicit sum-integral in Eq.~(\ref{eq:alternative}) is background independent, owing to the background symmetry.} see Eq.~(\ref{eq:Mvac}). As for the  thermal contribution, it reads $\hat M^2_{T,jk}T^2/g^2$, with
\beq
\hat M^2_{T,jk} & \,\hat{=}\, & g^2\sum_\kappa\kappa^j\kappa^k\left[\int_Q^T\left(\frac{m^2}{q^2}+2+\frac{q^2}{m^2}\right)\frac{1}{Q^2_\kappa+m^2}\right.\nonumber\\
& & \hspace{2.5cm}\left.-\,\int_Q^T\left(\frac{1}{2}+\frac{q^2}{m^2}\right)\frac{1}{Q^2_\kappa}\right]\!,\nonumber\\
\eeq
to which we should of course add the expression for $M^2_{T=0}\delta_{jk}$ obtained above, see Eq.~(\ref{eq:Mvac}). Combining the results, we find
\beq
\hat M^2_{T,jk} & = & \left[Z_a Z_{m^2}+\frac{3g^2N}{64\pi^2}\left(\frac{1}{\epsilon}+\ln\frac{\bar\Lambda^2}{m^2}+\frac{5}{6}\right)\right]m^2\delta_{jk}\nonumber\\
& + & \frac{g^2}{2\pi^2}\sum_\kappa\kappa^j\kappa^k\int_0^\infty dq\,q^2\,{\rm Re}\nonumber\\
& & \times\,\left[\left(\frac{m^2}{q^2}+2+\frac{q^2}{m^2}\right)\frac{n_{\varepsilon_q^\kappa}}{\varepsilon_q}-\left(\frac{1}{2}+\frac{q^2}{m^2}\right)\frac{n_{q^\kappa}}{q}\right].\nonumber\\
\eeq
Moreover, we write
\beq
\delta \hat V_{\bar r}(r) & = & \sum_\kappa\int_Q^T \!\Big[\hat L(\Delta r)-\hat L(0)-\Delta r^j \hat L'_j(0)\nonumber\\
& & \hspace{2.0cm}-\,\frac{\Delta r^j\Delta r^k}{2}\hat L''_{jk}(0)\Big],
\eeq
with
\beq
\hat L(\Delta r)=\frac{1}{2}\ln\left[1+\frac{m^2\bar Q_\kappa^2}{(Q_\kappa\cdot\bar Q_\kappa)^2}\right],
\eeq
as well as
\beq
\hat L(0) & = & \frac{1}{2}\ln\Big[\bar Q^2_\kappa+m^2\Big]-\frac{1}{2}\ln \bar Q^2_\kappa\,,\\
\hat L'_j(0) & = & T\kappa^j \left[\frac{\bar\omega_q^\kappa}{\bar Q_\kappa^2+m^2}-\frac{\bar\omega_q^\kappa}{\bar Q_\kappa^2}\right],\\
\eeq
and
\beq
\hat L''_{jk}(0) & = & T^2\kappa^j\kappa^k\left[ \frac{(\bar\omega_q^\kappa)^2}{\bar Q_\kappa^4}-2\frac{(\bar\omega_q^\kappa)^2}{(\bar Q_\kappa^2+m^2)^2}\right.\nonumber\\
& & \hspace{2.2cm}\left.+\,\frac{(\bar\omega_q^\kappa)^2}{\bar Q^2_\kappa(\bar Q^2_\kappa+m^2)}\right]\!.
\eeq
Then,
\beq
\hat L(\Delta r) & = & \frac{1}{2}\ln\frac{(q^2+M^2_{+,\kappa})(q^2+M^2_{-,\kappa})}{(q^2+M^2_{0,\kappa})^2}
\eeq
and
\beq
\hat L(0) & = & \frac{1}{2}\ln(q^2+\bar M^2_{\kappa})-\frac{1}{2}\ln (q^2+\bar M^2_{0,\kappa})\,,\\
\hat L'_j(0) & = & T\kappa^j \left[\frac{\bar\omega_q^\kappa}{q^2+\bar M^2_{\kappa}}-\frac{\bar\omega_q^\kappa}{q^2+\bar M^2_{0,\kappa}}\right]\!,\\
\hat L''_{jk}(0) & = & T^2\kappa^j\kappa^k\left[ \frac{(\bar\omega_q^\kappa)^2}{(q^2+\bar M^2_{0,\kappa})^2}-2\frac{(\bar\omega_q^\kappa)^2}{(q^2+\bar M^2_{\kappa})^2}\right.\nonumber\\
& & \hspace{1.6cm}\left.+\,\frac{(\bar\omega_q^\kappa)^2}{(q^2+\bar M^2_{0,\kappa})(q^2+\bar M^2_{\kappa})}\right]\!.\nonumber\\
\eeq
Putting all the pieces together, we arrive at
\begin{widetext}
\beq
& & \delta \hat V_{\bar r}(r)=T\sum_\kappa \sum_{q\in\mathds{Z}}\int\frac{d^Dq}{(2\pi)^D}\left[\frac{1}{2}\ln \frac{(q^2+M^2_{+,\kappa})(q^2+M^2_{-,\kappa})(q^2+\bar M^2_{0,\kappa})}{(q^2+M^2_{0,\kappa})^2(q^2+\bar M^2_\kappa)}\right.\nonumber\\
& & \hspace{4.5cm}+\,T(\kappa\cdot\Delta r)\left(\frac{\bar \omega_q^\kappa}{q^2+\bar M^2_{0,\kappa}}-\frac{\bar \omega_q^\kappa}{q^2+\bar M^2_\kappa}\right)\nonumber\\
& & \hspace{2.0cm}\left.-\,\frac{1}{2}T^2(\kappa\cdot\Delta r)^2\left(\frac{(\bar\omega_q^\kappa)^2}{(q^2+\bar M^2_{0,\kappa})^2}+\frac{(\bar \omega_q^\kappa)^2}{(q^2+\bar M^2_{0,\kappa})(q^2+\bar M^2_\kappa)}-2\frac{(\bar\omega_q^\kappa)^2}{(q^2+\bar M^2_\kappa)^2}\right)\right]\!.
\eeq
Performing the $D$-dimensional integrals, one arrives this time at
\beq
& &  \delta \hat V_{\bar r}(r)=\frac{T}{(4\pi)^{D/2}}\sum_\kappa\sum_{q\in\mathds{Z}}\Bigg[\frac{1}{2}\Big(2\,{\rm Re}\,(M^2_{0,\kappa})^{D/2}+(\bar M^2_{\kappa})^{D/2}-(\bar M^2_{0,\kappa})^{D/2}-(M^2_{\kappa,+})^{D/2}-(M^2_{\kappa,-})^{D/2}\Big)\Gamma\left(-D/2\right)\nonumber\\
& & \hspace{4.5cm} +\,T(\kappa\cdot\Delta r)\,\bar\omega_q^\kappa\Big((\bar M^2_{0,\kappa})^{D/2-1}-(\bar M^2_\kappa)^{D/2-1}\Big)\Gamma\left(1-D/2\right)\nonumber\\
& & \hspace{4.5cm} +\,\frac{1}{2}T^2(\kappa\cdot\Delta r)^2(\bar\omega_q^\kappa)^2\Big(2(\bar M^2_\kappa)^{D/2-2}-(\bar M^2_{0,\kappa})^{D/2-2}\Big)\Gamma\left(2-D/2\right)\nonumber\\
& & \hspace{4.5cm} +\,\frac{1}{2}T^2(\kappa\cdot\Delta r)^2(\bar\omega_q^\kappa)^2\frac{(\bar M^2_\kappa)^{D/2-1}-(\bar M^2_{0,\kappa})^{D/2-1}}{\bar M^2_{\kappa}-\bar M^2_{0,\kappa}}\Gamma\left(1-D/2\right)\Bigg].
\label{eq:final_sum2}
\eeq
\end{widetext}
It can now be checked that the summand decreases at least as $1/\omega_q^2$ as $|\omega_q|\to\infty$. It follows that the $q$-summation and the $\epsilon\to 0$ limit can be permuted without missing any finite parts. One finally arrives at
\begin{widetext}
\beq
\hat V_{\bar r}(r) & = & \hat V_{\bar r}(\bar r)+\frac{T}{2\pi^2}\sum_\kappa (\kappa\cdot\Delta r) \int_0^\infty \!\!\!dq\,q^2\,{\rm Im}\big[n_{\varepsilon^\kappa_q}-n_{q^\kappa}\big]+\frac{T^2}{2g^2}\hat M^2_{T,jk} \Delta r^j\Delta r^k\nonumber\\
& + & \frac{T}{\pi}\sum_\kappa\sum_{q\in\mathds{Z}}\Bigg[\frac{1}{12}\Big(2\,{\rm Re}\,(M^2_{0,\kappa})^{3/2}+(\bar M^2_{\kappa})^{3/2}-(\bar M^2_{0,\kappa})^{3/2}-(M^2_{\kappa,+})^{3/2}-(M^2_{\kappa,-})^{3/2}\Big)\nonumber\\
& & \hspace{2.5cm} -\,\frac{T}{4}(\kappa\cdot\Delta r)\,\bar\omega_q^\kappa\Big((\bar M^2_{0,\kappa})^{1/2}-(\bar M^2_\kappa)^{1/2}\Big)\nonumber\\
& & \hspace{2.5cm} +\,\frac{T^2}{16}(\kappa\cdot\Delta r)^2(\bar\omega_q^\kappa)^2\left(\frac{2}{(\bar M^2_\kappa)^{1/2}}-\frac{1}{(\bar M^2_{0,\kappa})^{1/2}}-\frac{2}{(\bar M^2_{0,\kappa})^{1/2}+(\bar M^2_\kappa)^{1/2}}\right)\Bigg].
\eeq
We have checked that this formula gives the same results as the formula (\ref{eq:V_final}) obtained in the main text.\\

\end{widetext}

\section{Matsubara sum-integrals}\label{app:sums}
Let us gather here discuss the evaluation of the Matsubara sum-integrals.

\subsection{Contour deformation}
In the simplest cases, one can perform the Matsubara sum first, using the contour deformation technique. Indeed, consider a sum
\beq
F\equiv T\sum_{n\in\mathds{Z}}f(i\omega_n)\,,\label{eq:F}
\eeq
with $\smash{\omega_n=2\pi n T}$ and $f(z)$ a complex function with simple poles $z_i$ (distinct from the $i\omega_n$ for the sum to make sense) and such that $\smash{f(|z|)\to 0}$ fast enough as $\smash{|z|\to\infty}$. Introducing the complex version of the Bose-Einstein distribution function $n(z)\equiv 1/(e^{\beta z}-1)$, one can consider the contour integral
\beq
I_N\equiv \int_{C_N}\frac{dz}{2\pi i} f(z)n(z)\,,
\eeq
with $C_N$ a circle centered around $0$ with radius $\omega_{N+1/2}$. Since $n(z)$ has simple poles located precisely at the $i\omega_n$'s and because the corresponding residues are all equal to $1/\beta$, one finds from the residue theorem:
\beq
I_N=T\!\!\sum_{n=-N}^N f(i\omega_n)+\sum_{z_i\in D_n} {\rm Res}\,f|_{z_i}\,n(z_i)\,,
\eeq
where $D_N$ denotes the disk delimited by $C_N$. In the limit $N\to 0$, $\smash{I_N\to 0}$ due to the rapid vanishing of $f(z)$ as $\smash{|z|\to\infty}$. It follows that
\beq
F=-\sum_i {\rm Res}\,f|_{z_i}\,n(z_i)\,,\label{eq:F_result}
\eeq
which is tractable so long as one can easily identify the poles $z_i$ of $f(z)$.

In the main text, we need to evaluate sum-integrals of the form
\beq
\int_Q^T \frac{X(q)}{Q_\kappa^2+m^2}\,.
\eeq
Since
\beq
\frac{1}{Q_\kappa^2+m^2} & = & \frac{1}{(\omega^\kappa_q)^2+\varepsilon_q^2}\nonumber\\
& = & \frac{1}{2\varepsilon_q}\left[-\frac{1}{i\omega^\kappa_q-\varepsilon_q}+\frac{1}{i\omega^\kappa_q+\varepsilon_q}\right],
\eeq
the searched-after poles are $\pm\varepsilon_q-iTr\cdot \kappa$ with residues $\mp 1/(2\varepsilon_q)$. According to the general formula derived above, it follows that
\beq
\int_Q^T \frac{X(q)}{Q_\kappa^2+m^2}=\int\frac{d^Dq}{(2\pi)^D}X(q)\frac{n_{\varepsilon_q-iTr\cdot \kappa}-n_{-\varepsilon_q-iTr\cdot \kappa}}{2\varepsilon_q}\,.\nonumber\\
\eeq
Using $n_{-x}=-1-n_x$, this formula becomes
\beq
\int_Q^T \frac{X(q)}{Q_\kappa^2+m^2} & = & \int\frac{d^Dq}{(2\pi)^D}\frac{X(q)}{2\varepsilon_q}+\int\frac{d^Dq}{(2\pi)^D}X(q)\frac{{\rm Re}\,n_{\varepsilon^\kappa_q}}{\varepsilon_q}\,,\nonumber\\
\eeq
where we have introduced $\smash{\varepsilon^\kappa_q\equiv \varepsilon_q-iTr\cdot \kappa}$.

We have also split the result into a UV finite thermal piece which approaches $0$ as $\smash{T\to 0}$ and a vacuum piece which contains the potential UV divergences depending on the asymptotic behavior of $X(q)$. This vacuum piece can actually be given a more covariant form. For instance, in the case $\smash{X(q)=1}$, one has
\beq
\int_Q^T \frac{1}{Q_\kappa^2+m^2} & = & \int\frac{d^dQ}{(2\pi)^d}\frac{1}{Q^2+m^2}+\int\frac{d^Dq}{(2\pi)^D}\frac{{\rm Re}\,n_{\varepsilon^\kappa_q}}{\varepsilon_q}\,.\label{eq:tad}\nonumber\\
\eeq
This is also possible in the case where $X(q)$ is a polynomial in $q^2$ since each power of $q^2$ can, in the vacuum piece, be replaced by the same power of $Q^2$ times an appropriate $d$-dependent factor. This type of rewriting allows one to easily extract the UV divergent pieces from well-known formulas for Feynman integrals in the vacuum. As for the thermal pieces, because they are finite, they can be evaluated numerically directly in $\smash{D=3}$ dimensions. We shall see other examples below where the vacuum/thermal splitting is more subtle.

Another type of sum-integral is
\beq
\int_Q^T \frac{X(q)\omega_q^\kappa}{Q_\kappa^2+m^2}\,.
\eeq
To make the Matsubara sum absolutely convergent, it is convenient to add $0$ as
\beq
\int_Q^T X(q)\left[\frac{\omega_q^\kappa}{Q_\kappa^2+m^2}-\frac{\omega_q}{Q^2+m^2}\right].
\eeq
Then, we write
\beq
& & \frac{\omega_q^\kappa}{Q_\kappa^2+m^2}-\frac{\omega_q}{Q^2+m^2}\nonumber\\
& & \hspace{0.5cm}\,=\frac{1}{2i}\left[\frac{1}{i\omega_q-\varepsilon_q}+\frac{1}{i\omega_q+\varepsilon_q}\right.\nonumber\\
& & \hspace{1.5cm}\left.-\,\frac{1}{i\omega^\kappa_q-\varepsilon_q}-\frac{1}{i\omega^\kappa_q+\varepsilon_q}\right].
\eeq
The general formula (\ref{eq:F_result}) can now be applied and one finds
\beq
& & \int_Q^T \frac{X(q)\omega_q^\kappa}{Q_\kappa^2+m^2}\nonumber\\
& & =\,\int\frac{d^Dq}{(2\pi)^D}X(q)\frac{n_{\varepsilon_q-iTr\cdot \kappa}+n_{-\varepsilon_q-iTr\cdot \kappa}-n_{\varepsilon_q}-n_{-\varepsilon_q}}{2i}\nonumber\\
& & =\,\int\frac{d^Dq}{(2\pi)^D}X(q)\frac{n_{\varepsilon_q-iTr\cdot \kappa}+n_{-\varepsilon_q-iTr\cdot \kappa}+1}{2i}\nonumber\\
& & =\,\int\frac{d^Dq}{(2\pi)^D}X(q)\frac{n_{\varepsilon_q-iTr\cdot \kappa}-n_{\varepsilon_q=iTr\cdot \kappa}}{2i}\nonumber\\
& & =\,\int\frac{d^Dq}{(2\pi)^D}X(q)\,{\rm Im}\,n_{\varepsilon_q-iTr\cdot \kappa}\,.
\eeq
Notice that there is no vacuum contribution in this case.

\subsection{Momentum integration}
As we have discussed in the main text, another possible strategy is to perform the $D$-momentum integrals analytically and the resulting Matsubara sums numerically (after dealing with possible UV divergences). This is quite useful in cases where it is not simple to identify the poles $z_i$ in the contour deformation technique. Let us here illustrate this alternative technique on a simple example that can actually be treated with both approaches. 

Consider the tadpole sum-integral in the LHS of Eq.~(\ref{eq:tad}). The corresponding $D$-momentum integral is a vacuum tadpole integral of mass $M^2\equiv m^2+(\omega^\kappa_q)^2$. Then
\beq
\int_Q^T \!\!\frac{1}{Q_\kappa^2+m^2}=\frac{\Gamma(1-D/2)}{(4\pi)^{D/2}}T\sum_{q\in\mathds{Z}}(m^2+(\omega^\kappa_q)^2)^{D/2-1}\,.\nonumber\\
\eeq
It may seem that one can take the limit $\smash{D\to 3}$ safely since $\Gamma(-1/2)$ is finite. However, the resulting Matsubara sum is divergent because the summand grows like $|\omega_q|$. This simply means that $D$ plays the role of a regulator for the sum and before evaluating the latter numerically, one needs to extract the corresponding divergent part. This amounts to adding and subtracting the asymptotic contributions of the summand that either grow or do not decrease fast enough:
\begin{widetext}
\beq
\int_Q^T \frac{1}{Q_\kappa^2+m^2} & = & \frac{\Gamma(1-D/2)}{(4\pi)^{D/2}}T\sum_{q\in\mathds{Z}}\left[(m^2+(\omega^\kappa_q)^2)^{D/2-1}-|\omega^\kappa_q|^{D-2}-\left(\frac{D}{2}-1\right)m^2|\omega^\kappa_q|^{D-4}\right]\nonumber\\
& + & \frac{\Gamma(1-D/2)}{(4\pi)^{D/2}}T\sum_{q\in\mathds{Z}}|\omega^\kappa_q|^{D-2}-\frac{\Gamma(2-D/2)}{(4\pi)^{D/2}}m^2 T\sum_{q\in\mathds{Z}}|\omega^\kappa_q|^{D-4}\,.\label{eq:C10}
\eeq
\beq
(m^2+r^2+2r\omega_q+\omega_q^2)^{D/2-1} & = & |\omega_q|^{D-2}\left(1+\frac{2r}{\omega_q}+\frac{m^2+r^2}{\omega^2_q}\right)^{D/2-1}\nonumber\\
& = & |\omega_q|^{D-2}\left(1+(D-2)\frac{r}{\omega_q}+(D/2-1)\frac{m^2+r^2}{\omega^2_q}+2(D/2-1)(D/2-2)\frac{r^2}{\omega_q^2}\right)\nonumber\\
\eeq
\end{widetext}
In the subtracted sum (first line), one can safely take the limit $\smash{D\to 3}$ since it decreases as $1/|\omega^\kappa_q|^3$. On the other hand, the added terms (second line) are all of the form $\sum_{q\in\mathds{Z}}|\omega^\kappa_q|^{-s}$ which can be expressed in terms of the Hurwitz zeta function:
\beq
\zeta(s,z)\equiv\sum_{q=0}^\infty \frac{1}{(q+z)^s}\,.
\eeq
Indeed, after multiplying by the appropriate factors of $2\pi T$, we have
\beq
\sum_{q\in\mathds{Z}}^\infty \frac{(2\pi T)^s}{|\omega^\kappa_q|^s}=\sum_{q\in\mathds{Z}}^\infty \frac{1}{\left|q+\frac{\kappa\cdot r}{2\pi}\right|^s}\,.
\eeq
Since we can shift the summation variable $q$ by any integer $k$, we can assume replace $\kappa\cdot r/2\pi$ by $\{\kappa\cdot r/2\pi\}$ defined as the real number between $0$ and $1$ such that $\kappa\cdot r/2\pi-\{\kappa\cdot r/2\pi\}$ is an integer. We then have
\beq
\sum_{q\in\mathds{Z}}^\infty \frac{(2\pi T)^s}{|\omega^\kappa_q|^s} & = & \sum_{q=0}^\infty \frac{1}{\left(q+\left\{\frac{\kappa\cdot r}{2\pi}\right\}\right)^s}+\sum_{q=-\infty}^{-1} \frac{1}{\left(-q-\left\{\frac{\kappa\cdot r}{2\pi}\right\}\right)^s}\nonumber\\
& = & \sum_{q=0}^\infty \frac{1}{\left(q+\left\{\frac{\kappa\cdot r}{2\pi}\right\}\right)^s}+\sum_{q=0}^\infty \frac{1}{\left(q+1-\left\{\frac{\kappa\cdot r}{2\pi}\right\}\right)^s}\nonumber\\
& = & \zeta\left(s,\left\{\frac{\kappa\cdot r}{2\pi}\right\}\right)+\zeta\left(s,1-\left\{\frac{\kappa\cdot r}{2\pi}\right\}\right)\!.\nonumber\\
\eeq
Going back to Eq.~(\ref{eq:C10}), we need this formula for $\smash{s=2-D=-1+2\epsilon}$ or $\smash{s=4-D=1+2\epsilon}$, with $\smash{\epsilon\to 0}$. We can use
\beq
\zeta(-n,z)=-\frac{B_{n+1}(z)}{n+1}\,,
\eeq
where $B_n(z)$ denotes the Bernouilli polynomial of order $n$, and
\beq
\zeta(1+2\epsilon,z)=\frac{1}{2\epsilon}-\psi(z)\,,
\eeq
where $\smash{\psi(z)\equiv\Gamma'(z)/\Gamma(z)}$ denotes the digamma function. Putting all the pieces together, we finally arrive at
\begin{widetext}
\beq
\int_Q^T \frac{1}{Q_\kappa^2+m^2} & = & -\frac{T}{4\pi}\sum_{q\in\mathds{Z}}\left[(m^2+(\omega^\kappa_q)^2)^{1/2}-|\omega^\kappa_q|-\frac{m^2}{2|\omega^\kappa_q|}\right]+\frac{T^2}{2} B_2\left(\left\{\frac{r\cdot\kappa}{2\pi}\right\}\right)\nonumber\\
&  & -\,\frac{m^2}{16\pi^2}\left[\frac{1}{\epsilon}+\ln\frac{\bar\Lambda^2}{(4\pi T)^2}-\psi\left(\left\{\frac{r\cdot\kappa}{2\pi}\right\}\right)-\psi\left(1-\left\{\frac{r\cdot\kappa}{2\pi}\right\}\right)\right],\label{eq:tad_res}
\eeq
\end{widetext}
where we have used that $\smash{\psi(1/2)=-2\ln 2-\gamma}$, as well as $\smash{B_2(1-x)=B_2(x)}$. We have checked numerically that the right-hand sides of Eqs.~(\ref{eq:tad}) and (\ref{eq:tad_res}) coincide.

\subsection{The case of massless integrals}
By taking the limit $m\to 0$ in the formula above, we obtain
\beq
\int_Q^T \frac{1}{Q_\kappa^2} & = &  \frac{T^2}{2}B_2\left(\left\{\frac{r\cdot\kappa}{2\pi}\right\}\right),\label{eq:tad0}
\eeq
which provides an analytical expression for (\ref{eq:tad}) in the limit $\smash{m\to 0}$. We notice that there is no vacuum contribution in 
Eq.~(\ref{eq:tad0}), and, therefore, no divergence. This is in line with the fact that, in the limit $\smash{m\to 0}$, the vacuum contribution to (\ref{eq:tad}) becomes a scaleless integral which vanishes in dimensional regularization.

Similarly, by acting on Eq.~(\ref{eq:tad_res}) with the operator $-d/dm^2$ and then taking the limit $\smash{m\to 0}$, one obtains
\beq
\int_Q^T \frac{1}{Q_\kappa^4} & = & \frac{1}{16\pi^2}\Bigg[\frac{1}{\epsilon}+\ln\frac{\bar\Lambda^2}{(4\pi T)^2}\\
& & -\,\psi\left(\left\{\frac{r\cdot\kappa}{2\pi}\right\}\right)-\psi\left(1-\left\{\frac{r\cdot\kappa}{2\pi}\right\}\right)\Bigg],\nonumber
\eeq
where $Q_\kappa^4$ is a short-hand notation for $(Q^2_\kappa)^2$. We notice that there is a pole in $1/\epsilon$ in this case. This seems in contradiction with the thermal spliting that one would derive from Eq.~(\ref{eq:tad})
\beq
\int_Q^T \!\frac{1}{Q_\kappa^4}\!=\!\!\int\!\frac{d^Dq}{(2\pi)^D}\frac{1}{2q}\frac{d}{dq}\frac{1}{2q}+\!\int\!\frac{d^Dq}{(2\pi)^D}\frac{1}{2q}\frac{d}{dq}\frac{{\rm Re}\,n_{q-iTr\cdot\kappa}}{q}\,,\nonumber\\
\eeq
and that seems to feature once more a vanishing scaleless integral in the vacuum contribution. The problem here is that the thermal splitting does not make any sense due to the presence of an IR divergence $d^Dq/q^3 {\rm Re}\,n_{q-iTr\cdot\kappa}$. In fact, after noticing that
\beq
{\rm Re}\,n_{-iTr\cdot\kappa} & = & {\rm Re}\,\frac{1}{e^{-ir\cdot\kappa}-1}\nonumber\\
& = & {\rm Re}\,\frac{e^{ir\cdot\kappa/2}}{e^{-ir\cdot\kappa/2}-e^{ir\cdot\kappa/2}}\nonumber\\
& = & -\frac{1}{2}{\rm Re}\,\frac{e^{ir\cdot\kappa/2}}{i\sin(r\cdot\kappa/2)}\nonumber\\
& = & -\frac{1}{2}{\rm Im}\,\frac{e^{ir\cdot\kappa/2}}{\sin(r\cdot\kappa/2)}=-\frac{1}{2}\,,
\eeq
one can rewrite the ``vacuum contribution as''
\beq
\int\frac{d^Dq}{(2\pi)^D}\frac{1}{2q}\frac{d}{dq}\frac{1}{2q}=-\int\frac{d^Dq}{(2\pi)^D}\frac{1}{2q}\frac{d}{dq}\frac{{\rm Re}\,n_{-iTr\cdot\kappa}}{q}\,,\nonumber\\
\eeq
which upon combination with the ``thermal piece'' gives
\beq
\int_Q^T \frac{1}{Q_\kappa^4} & = & \int\frac{d^Dq}{(2\pi)^D}\frac{1}{2q}\frac{d}{dq}\frac{{\rm Re}\,(n_{q-iTr\cdot\kappa}-n_{-iTr\cdot\kappa})}{q}\,,\nonumber\\
\eeq
which is now IR-safe. In conclusion, the vacuum piece cannot be separated from the thermal piece in this case since it contributes to rendering the original integral IR-safe.

Lastly, we note that we can also use Eq.~(\ref{eq:C10}) in order to evaluate
\beq
\int_Q^T \frac{q^2}{\bar Q^2_\kappa}=-\int_Q^T \frac{(\bar\omega^\kappa_q)^2}{\bar Q^2_\kappa}\,.
\eeq
Indeed, the only modification to be made in the RHS of Eq.~(\ref{eq:C10}) is an overall minus sign and an extra factor of $(\bar\omega^\kappa_q)^2$ in all the summands. In the limit $\smash{m\to 0}$, only the second sum survives, as in the case of $\int_Q^T 1/\bar Q^2_\kappa$. One eventually finds
\beq
\int_Q^T \frac{q^2}{\bar Q^2_\kappa}=-\pi^2T^2 B_4\left(\left\{\frac{\kappa\cdot\bar r}{2\pi}\right\}\right).\label{eq:tad4}
\eeq
We have used both Eq.~(\ref{eq:tad0}) and (\ref{eq:tad4}) in Eq.~(\ref{eq:curv4}).

\subsection{Some final remarks}
Let us go back to the general definition (\ref{eq:F}) and suppose we choose the summand to be $f_\kappa(z)\equiv f(z+iT\kappa\cdot r)$, so that the sum is now
\beq
F_\kappa(T)=T\sum_{n\in\mathds{Z}}f(i\omega_n+iT\kappa\cdot r)\,.
\eeq
Suppose also that we are actually interested in the sum $\sum_\kappa F_\kappa(T)$ in the case $\smash{r=\bar r_c}$. As we now show, the result can be easily expressed in terms of $F_{\kappa=0}(T)$.

First of all, we note that one possible choice of $\bar r_c$ is
\beq
\bar r_c=\frac{4\pi}{N}\sum_{k=1}^N (N-k+1)\rho_k\,,
\eeq
see for instance Ref.~\cite{vanEgmond:2023lfu}. Second, we note that $\kappa$ is either $0$, in which case $\smash{\kappa\cdot\bar r_c=0}$ or a difference $\rho_i-\rho_j$, in which case $\smash{\kappa\cdot\bar r_c=(2\pi/N)(j-i)}$. It follows that
\beq
\sum_\kappa F_\kappa(T) & = & (N-1)F_{\kappa=0}(T)\\
& + & T\sum_{i\neq j}\sum_{n\in\mathds{Z}}f\left(i2\pi \frac{T}{N}(Nn+j-i)\right).\nonumber
\eeq
It is convenient to split the sum as
\beq
\sum_\kappa F_\kappa(T) & = & (N-1)F_{\kappa=0}(T)\nonumber\\
& + & T\sum_{i=1}^N\sum_{j=1}^{i-1}\sum_{n\in\mathds{Z}}f\left(i2\pi \frac{T}{N}(Nn+j-i)\right)\nonumber\\
& + & T\sum_{i=1}^N\sum_{j=i+1}^N\sum_{n\in\mathds{Z}}f\left(i2\pi \frac{T}{N}(Nn+j-i)\right).\nonumber\\
\eeq
In the second line, we then do $n\to n+1$ and $j\to j-N$. Then
\beq
\sum_\kappa F_\kappa(T) & = & (N-1)F_{\kappa=0}(T)\nonumber\\
& + & T\sum_{i=1}^N\sum_{j=i+1}^{N+i-1}\sum_{n\in\mathds{Z}}f\left(i2\pi \frac{T}{N}(Nn+j-i)\right).\nonumber\\
\eeq
We can now do $j\to j+i$ and then
\beq
\sum_\kappa F_\kappa(T) & = & (N-1)F_{\kappa=0}(T)\nonumber\\
& + & TN\sum_{j=1}^{N-1}\sum_{n\in\mathds{Z}}f\left(i2\pi \frac{T}{N}(Nn+j)\right),\nonumber\\
\eeq
or
\beq
\sum_\kappa F_\kappa(T) & = & -F_{\kappa=0}(T)\nonumber\\
& + & N^2\frac{T}{N}\sum_{j=0}^{N-1}\sum_{n\in\mathds{Z}}f\left(i2\pi \frac{T}{N}(Nn+j)\right).\nonumber\\
\eeq
This is nothing but
\beq
\sum_\kappa F_\kappa(T) & = & -F_{\kappa=0}(T)+N^2F_{\kappa=0}(T/N)\,,
\eeq
which relates the looked-after sum of the various $F_\kappa(T)$'s to $F_{\kappa=0}(T)$.

 Using the same type of manipulations, let us show that $\smash{\sum_\kappa \kappa F_\kappa(T)=0}$ when $\smash{\bar r=\bar r_c}$. All the information in this sum is contained in the projections along the weights $\rho_k$
\beq
& & \sum_\kappa (\rho_k\cdot\kappa) F_\kappa(T)\nonumber\\
& & \hspace{0.5cm}=\,T\sum_{i\neq j}\sum_{n\in\mathds{Z}}\rho_k\cdot(\rho_i-\rho_j)f\left(i2\pi \frac{T}{N}(Nn+j-i)\right)\nonumber\\
& & \hspace{0.5cm}=\,\frac{T}{2}\sum_{j\neq k}\sum_{n\in\mathds{Z}}f\left(i2\pi \frac{T}{N}(Nn+j-k)\right)\nonumber\\
& & \hspace{0.5cm}-\,\frac{T}{2}\sum_{j\neq k}\sum_{n\in\mathds{Z}}f\left(i2\pi \frac{T}{N}(Nn+k-j)\right).
\eeq
Splitting the sums, we find
\beq
& & \sum_\kappa (\rho_k\cdot\kappa) F_\kappa(T)\nonumber\\
& & \hspace{0.5cm}=\,\frac{T}{2}\sum_{j=1}^{k-1}\sum_{n\in\mathds{Z}}f\left(i2\pi \frac{T}{N}(Nn+j-k)\right)\nonumber\\
& & \hspace{0.5cm}+\,\frac{T}{2}\sum_{j=k+1}^{N}\sum_{n\in\mathds{Z}}f\left(i2\pi \frac{T}{N}(Nn+j-k)\right)\nonumber\\
& & \hspace{0.5cm}-\,\frac{T}{2}\sum_{j=1}^{k-1}\sum_{n\in\mathds{Z}}f\left(i2\pi \frac{T}{N}(Nn+k-j)\right)\nonumber\\
& & \hspace{0.5cm}-\,\frac{T}{2}\sum_{j=k+1}^N\sum_{n\in\mathds{Z}}f\left(i2\pi \frac{T}{N}(Nn+k-j)\right).\nonumber\\
\eeq
Using the same trick as above, this rewrites
\beq
& & \sum_\kappa (\rho_k\cdot\kappa) F_\kappa(T)\nonumber\\
& & \hspace{0.5cm}=\,\frac{T}{2}\sum_{j=1}^{N-1}\sum_{n\in\mathds{Z}}f\left(i2\pi \frac{T}{N}(Nn+j)\right)\nonumber\\
& & \hspace{0.5cm}-\,\frac{T}{2}\sum_{j=1}^{N-1}\sum_{n\in\mathds{Z}}f\left(i2\pi \frac{T}{N}(Nn-j)\right),
\eeq
which is seen to vanish upon the change of variables $\smash{n\to n+1}$ and $\smash{j\to N-j}$ in the second sum.

\section{Matsubara sums and $\epsilon$-expansion}\label{app:eps}
We have already mentioned that, even though $\delta V_{\bar r}(r)$ admits a finite $\smash{\epsilon\to 0}$ limit, taking the limit in Eq.~(\ref{eq:final_sum}) is tricky because it does not commute with the Matsubara summation. The culprit is the presence of a contribution $\epsilon\times 1/|\omega_q|^{1+2\epsilon}$ in the asymptotic expansion of the summand at large $|\omega_q|$. This contribution vanishes if the limit is taken before the sum, but gives a finite non-zero contribution if the limit is taken after the sum. 

In the main text, we have subtracted the problematic terms from the summand and we have added them back in the form of a series that can be expressed in terms of the Hurwitz zeta function. Another possible strategy is to take the naive limit of the summand anyway and add the finite contribution by hand. This strategy was used in Ref.~\cite{vanEgmond:2022nuo} when evaluating the gluon propagator in the center-symmetric Landau gauge. In the present case, it leads to

\begin{widetext}
\beq
\delta V_{\bar r}(r) & = & \frac{T^4}{6\pi^2}\sum_\kappa(\kappa\cdot\Delta r)^3\kappa\cdot\left(\bar r+\frac{\Delta r}{4}\right)\nonumber\\
& + & \frac{T}{\pi}\sum_\kappa\sum_{q\in\mathds{Z}}\Bigg[\frac{1}{12}\Big(2(M^2_{0,\kappa})^{3/2}+3(\bar M^2_{\kappa})^{3/2}-(\bar M^2_{0,\kappa})^{3/2}-2(M^2_{\kappa})^{3/2}-(M^2_{\kappa,+})^{3/2}-(M^2_{\kappa,-})^{3/2}\Big)\nonumber\\
& & \hspace{2.5cm} -\,\frac{T}{4}(\kappa\cdot\Delta r)\,\bar\omega_q^\kappa\Big((\bar M^2_{0,\kappa})^{1/2}-3(\bar M^2_\kappa)^{1/2}\Big)+\frac{T^2}{4}(\kappa\cdot\Delta r)^2(\bar M^2_\kappa)^{1/2}\nonumber\\
& & \hspace{2.5cm} +\,\frac{T^2}{16}(\kappa\cdot\Delta r)^2(\bar\omega_q^\kappa)^2\left(\frac{6}{(\bar M^2_\kappa)^{1/2}}-\frac{1}{(\bar M^2_{0,\kappa})^{1/2}}-\frac{2}{(\bar M^2_{0,\kappa})^{1/2}+(\bar M^2_\kappa)^{1/2}}\right)\Bigg],\label{eq:alt}
\eeq
\end{widetext}
where the correction terms in the first line come from the second term of Eq.~(\ref{eq:asym}) which has precisely the form $\epsilon\times 1/|\omega_q|^{1+2\epsilon}$. We should stress, however, that Eq.~(\ref{eq:alt}) hides a subtle point which we now discuss.

Indeed, let us write Eq.~(\ref{eq:final_sum}) as
\beq
\delta V_{\bar r}(r) & = & \frac{T^4}{6\pi}\sum_\kappa(\kappa\cdot\Delta r)^3\left(2\left\{\frac{\kappa\cdot\bar r}{2\pi}\right\}-1+\frac{\kappa\cdot\Delta r}{4\pi}\right)\nonumber\\
& + & \frac{T}{\pi}\sum_\kappa\sum_{q\in\mathds{Z}}\Bigg[s_q+\frac{T^3}{6}(\kappa\cdot\Delta r)^3{\rm sgn}(\bar\omega^\kappa_q)\Bigg].\label{eq:final_sum_22}
\eeq
The sum in this case is absolutely convergent which means that it should not depend on the way it is computed. Suppose for instance that we evaluate it as $\lim_{N\to\infty}\sum_{q=-N}^{N+n}$, where $n$ has been introduced for convenience but should not affect the final result. Then we have
\beq
\delta V_{\bar r}(r) & = & \frac{T^4}{6\pi}\sum_\kappa(\kappa\cdot\Delta r)^3\left(2\left\{\frac{\kappa\cdot\bar r}{2\pi}\right\}-1+\frac{\kappa\cdot\Delta r}{4\pi}\right)\nonumber\\
& + & \frac{T}{\pi}\sum_\kappa\lim_{N\to\infty}\sum_{q=-N}^{N+n}\Bigg[s_q+\frac{T^3}{6}(\kappa\cdot\Delta r)^3{\rm sgn}(\bar\omega^\kappa_q)\Bigg].\nonumber\\
\eeq
We now use
\beq
& & \sum_{q=-N}^{N+n}{\rm sgn}\left(q+\frac{\kappa\cdot\bar r}{2\pi}\right)\nonumber\\
& & \hspace{0.5cm}=\,\sum_{q=-N+\frac{\kappa\cdot\bar r}{2\pi}-\left\{\frac{\kappa\cdot\bar r}{2\pi}\right\}}^{N+n+\frac{\kappa\cdot\bar r}{2\pi}-\left\{\frac{\kappa\cdot\bar r}{2\pi}\right\}}{\rm sgn}\left(q+\left\{\frac{\kappa\cdot\bar r}{2\pi}\right\}\right)\nonumber\\
& & \hspace{0.5cm}=\,n+2\left(\frac{\kappa\cdot\bar r}{2\pi}-\left\{\frac{\kappa\cdot\bar r}{2\pi}\right\}\right)+1\,,
\eeq
and then
\beq
\delta V_{\bar r}(r) & = & \frac{T^4}{6\pi}\sum_\kappa(\kappa\cdot\Delta r)^3\left(n+2\frac{\kappa\cdot\bar r}{2\pi}+\frac{\kappa\cdot\Delta r}{4\pi}\right)\nonumber\\
& + & \frac{T}{\pi}\sum_\kappa\lim_{N\to\infty}\sum_{q=-N}^{N+n} s_q\,.\label{eq:alt2}
\eeq
We recognize Eq.~(\ref{eq:alt}) provided we choose $\smash{n=0}$. This means that Eq.~(\ref{eq:alt}) is actually valid only provided one chooses the symmetric summation $\lim_{N\to\infty}\sum_{q=-N}^N$.\footnote{This is the summation that was used in Ref.~\cite{vanEgmond:2022nuo}.} For non-symmetric summations, the correction term needs to be modified according to Eq.~(\ref{eq:alt2}). This subtlety originates in the fact that, unlike the sum in Eq.~(\ref{eq:final_sum}), the sum in Eq.~(\ref{eq:alt}), although convergent, is not absolutely convergent.

A related aspect is that, in the case of the formula (\ref{eq:alt2}), some of the symmetries discussed in Sec.~\ref{sec:sym} are only manifest once one adds all the terms. This is for instance the case of the symmetry $\smash{(r,\bar r)\to (r+4\pi\alpha,\bar r+4\pi\alpha)}$. Under such transformation, the first term of Eq.~(\ref{eq:alt2}) is shifted by
\beq
\frac{2T^4}{3\pi}\sum_\kappa(\kappa\cdot\Delta r)^3\kappa\cdot\alpha\,.\label{eq:shift}
\eeq
On the other hand, the second term becomes
\beq
\frac{T}{\pi}\sum_\kappa\lim_{N\to\infty}\sum_{q=-N}^{N+n} s_{q+2\kappa\cdot\alpha}\,,
\eeq
where we have used the fact that $s_q$ depends on $r$ and $\bar r$ only via $\omega_q^\kappa$ and $\bar\omega_q^\kappa$, and we recall that $2\kappa\cdot\alpha$ is always an integer. This rewrites
\beq
\frac{T}{\pi}\sum_\kappa\lim_{N\to\infty}\sum_{q=-N+2\kappa\cdot\alpha}^{N+n+2\kappa\cdot\alpha} s_{q}\,,
\eeq
which represents a shift of the corresponding term in Eq.~(\ref{eq:alt2}) by
\beq
\frac{T}{\pi}\sum_\kappa\lim_{N\to\infty}\left[\sum_{q=N+n+1}^{N+n+2\kappa\cdot\alpha} s_{q}-\sum_{q=-N}^{-N-1+2\kappa\cdot\alpha} s_{q}\right]\,.
\eeq
For large enough $N$, $s_q$ in each sum can be replaced by the constant $\mp T^4(\kappa\cdot\Delta r)^3/(6\pi)$ that originates from the first term of (\ref{eq:asym}) for $\smash{D=3}$. Since each sum counts $2\kappa\cdot\alpha$ terms, the total shift is
\beq
-\frac{2T^4(\kappa\cdot\Delta r)^3}{3\pi}
\eeq
which cancels identically the one, see Eq.~(\ref{eq:shift}), from the first term of Eq.~(\ref{eq:alt2}). In the case of Eq.~(\ref{eq:final_sum_22}), in contrast, each term is symmetric, due to the presence of $\{\kappa\cdot \bar r\}$ in the first term and the fact that the sum in the second term is absolutely convergent.

\end{document}